\documentclass[slac_one]{revtex4}
\usepackage{graphicx}
\usepackage{epstopdf}
\usepackage{fancyhdr}
\usepackage{xspace}
\usepackage{subfigure}
\usepackage{lineno}
\newcommand{\pT}{$p_T$\xspace}
\newcommand{\FigRef}[1]{Figure~\ref{#1}\xspace}

\begin{document}

\title{Minimum Bias and Early QCD at ATLAS} %% Paper title goes here

\author{D. Kar (on behalf of the ATLAS collaboration)}
\affiliation{IKTP, TU Dresden, Germany}

%\begin{abstract}
%The early Minimum bias and the QCD results from the ATLAS experiment in the LHC are presented.
%\end{abstract}

\maketitle

\thispagestyle{fancy}


\section{INTRODUCTION} % Section title should be in all capitals.

LHC~\cite{lhc} is primarily a discovery machine, and the main aim is to search for the Higgs boson, and generally for the signatures of physics beyond the standard model (SM). However, before claiming any such discovery, all the SM cross-sections, which are essential backgrounds to searches, need to be measured precisely. In particular, QCD is the essential ingredient, and most of the interesting physics signatures at involve final states with jets of hadrons. A larger phase space for gluon emission and thus for production of extra jets compared to Tevatron leads to more intensive QCD backgrounds at the LHC. So it is essential to understand the QCD processes, and to make sure QCD Monte-Carlo (MC) models give sensible results.
In general, exploring the different aspects of QCD in a new high energy and high multiplicity regime is by itself interesting.

\section{SOFT QCD MEASUREMENTS}

%%%Soft QCD

The processes of interest at hadron colliders are mostly the hard scattering events.
However, soft QCD processes are unavoidable background to all the collider observables and they are generally not well understood since non-perturbative physics is involved. So the soft QCD distributions are used to test the predictions from the phenomenological models implemented in various MC generators. 
%By tuning the knobs made available in generators, we try to match the simulation to the data in the best possible way in order to gain deeper insights into the the various contributing sub-processes. 
All the measurements in this section were performed with the ATLAS~\cite{atlas} detector for $\sqrt{s} =$ 0.9 and 7 TeV (Minimum bias results were also obtained at $\sqrt{s} =$ 2.36 TeV~\cite{mb_236}), and while only 7 TeV distributions are shown, the 0.9 TeV results are essentially similar.

%%%MinBias

\begin{figure}[htbp]
\centering
\subfigure[]{
\includegraphics[scale=0.18]{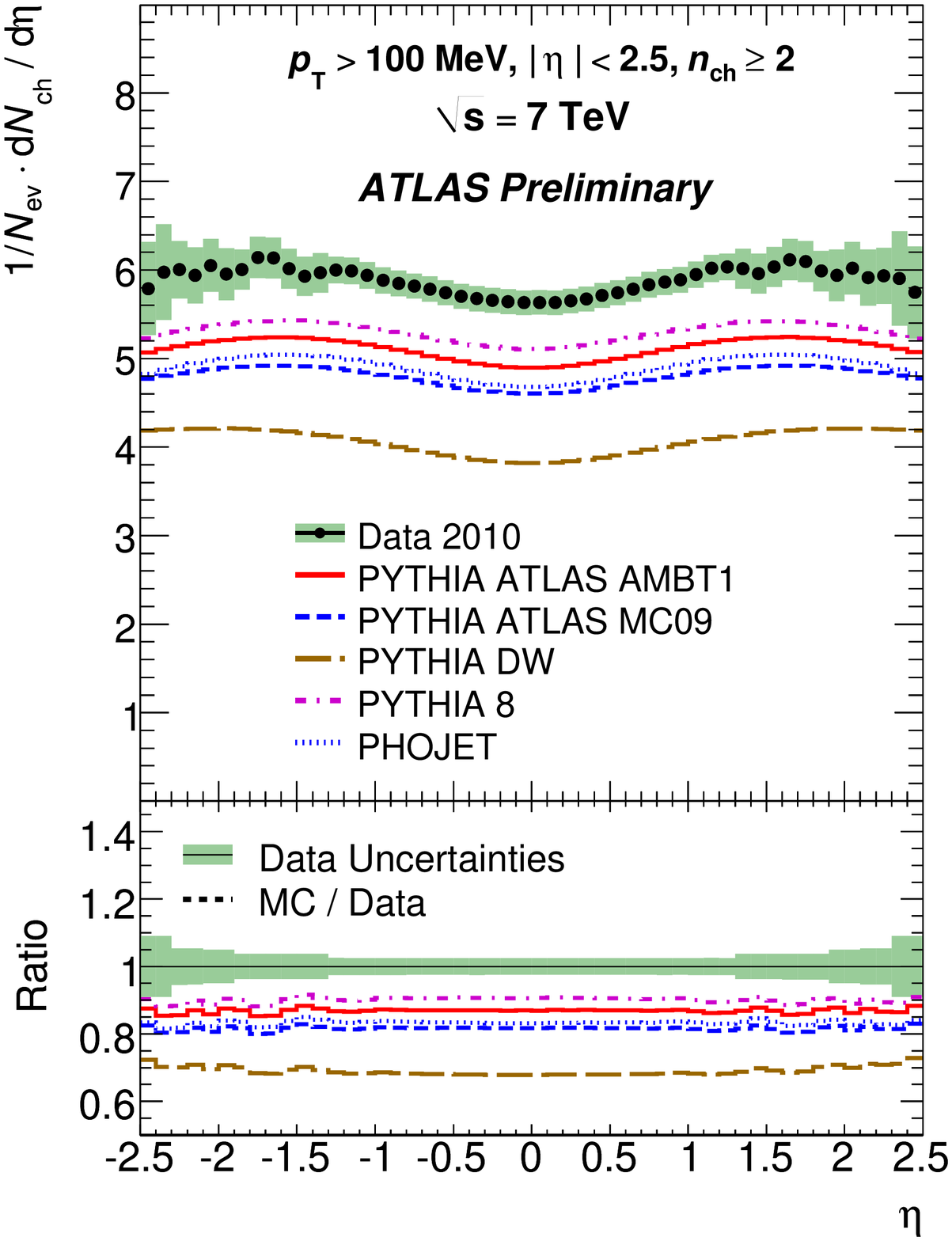}
%\label{fig:mb1}
}
\subfigure[]{
\includegraphics[scale=0.18]{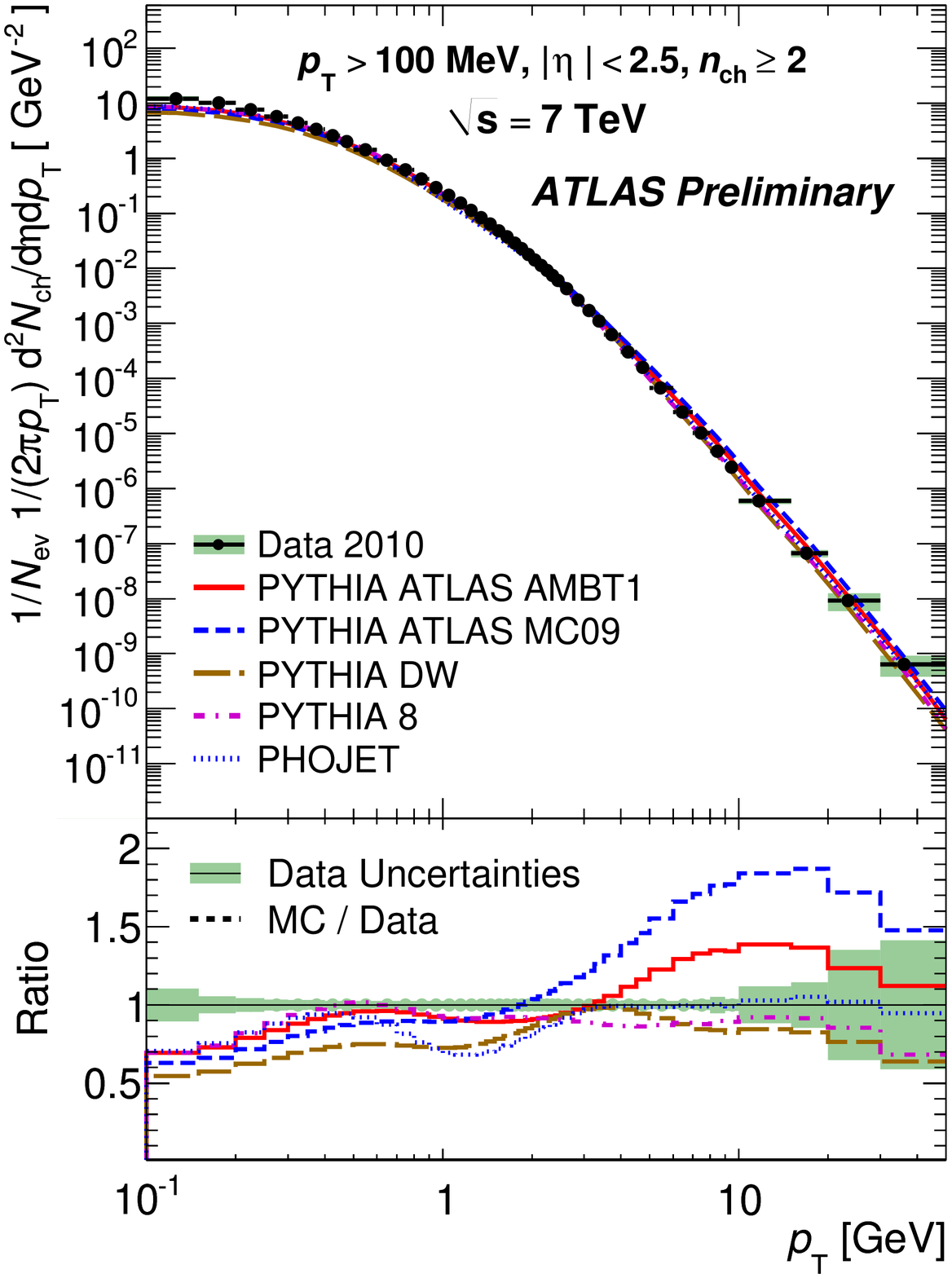}
%\label{fig:mb2}
}
\subfigure[]{
\includegraphics[scale=0.18]{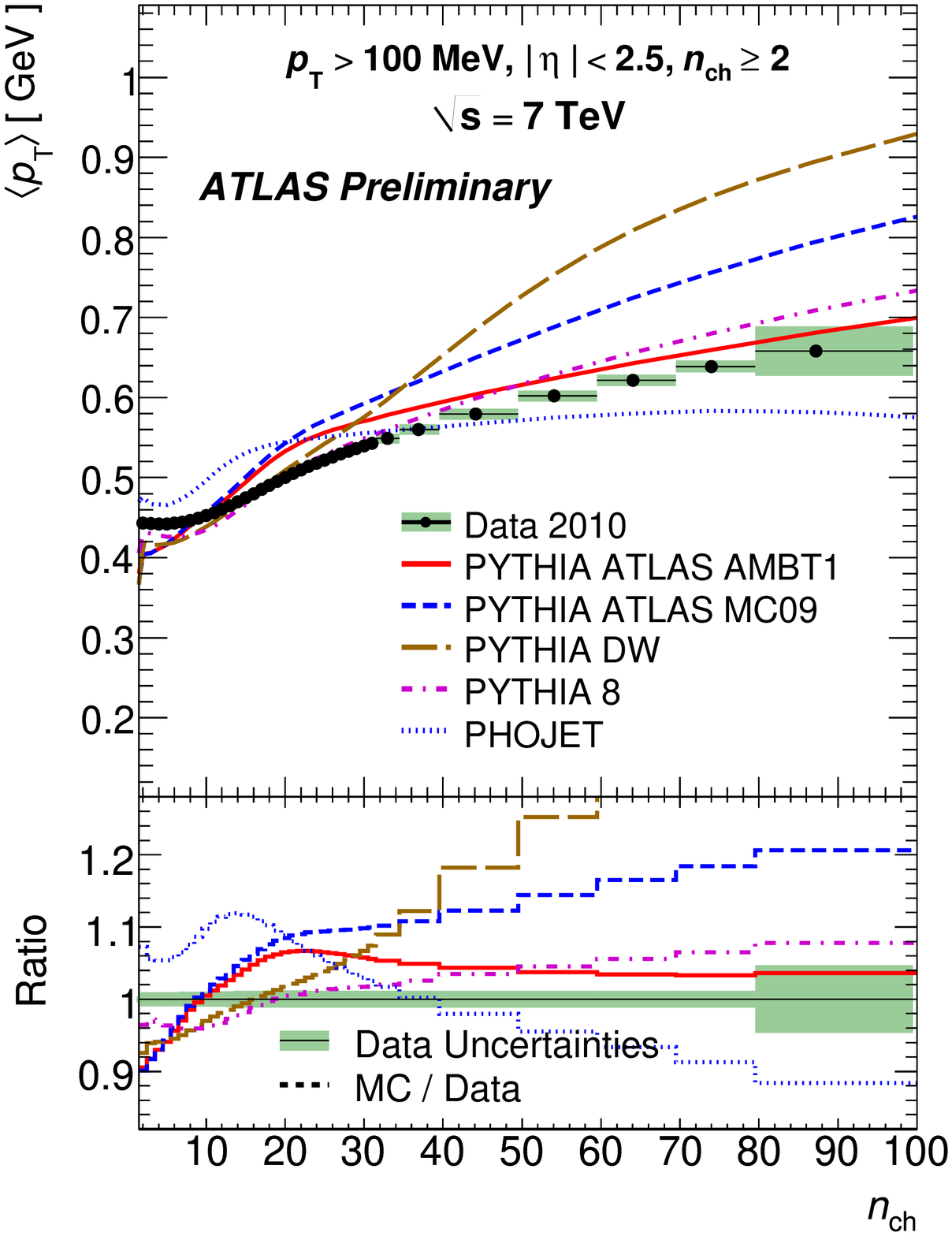}
%\label{fig:mb4}
}
\subfigure[]{
\includegraphics[scale=0.18]{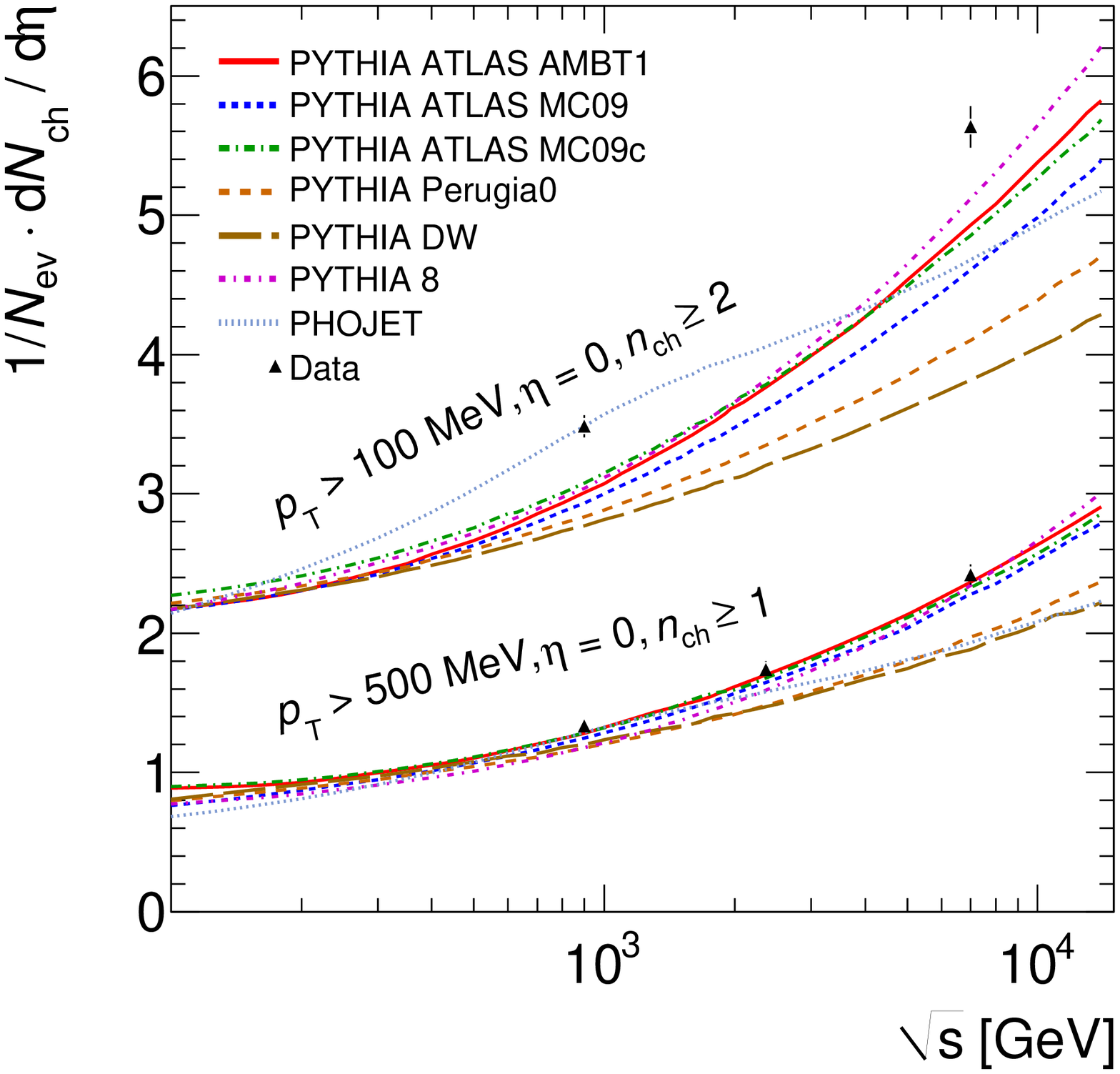}
%\label{fig:mb3}
}
\caption[]{Charged particle multiplicity is shown as a function of the pseudorapidity in (a) and transverse momentum in (b). The average transverse momentum as a function of the number of charged particles in the event is shown in (c). 
%The dots represent the data and the curves the predictions from different Monte Carlo models. 
%The vertical bars represent the statistical uncertainties, while the shaded areas show statistical and systematic uncertainties added in quadrature. The values of the ratio histograms refer to the bin centroids. 
In (d), the average charged particle multiplicity per unit of rapidity for $\eta = 0$ is plotted as a function of the center of mass energy and compared to various particle level MC predictions.}
\label{fig:mb} 
\end{figure}

Minimum bias is a generic term which refers to events that are selected with a 'loose' trigger that accepts a
large fraction of the inelastic cross-section (ideally with totally inclusive trigger).
In ATLAS, data were selected by a Minimum Bias Trigger Scintillator (MBTS) single-arm trigger. 
The data were corrected back to particle level by applying efficiency corrections and unfolding to account for migration and
compared to different MC models as fully inclusive-inelastic distributions with no model dependent corrections. 
The measurements were first performed with a \pT ~cut of 500 MeV~\cite{mb10}  
and subsequently low \pT ~tracking algorithm used to go down to 100 MeV~\cite{mb20}. 
Also diffraction suppressed and enhanced samples were looked into for testing and tuning MC models~\cite{mb_diff}.
The distributions for events with $N_{ch} >= 2$ within the kinematic range 
$p_T >= 100$ MeV and $|\eta| < 2.5$ are shown in \FigRef{fig:mb},
where $N_{ev}$ is the number of events with at least two charged particles inside the selected kinematic range,
$N_{ch}$ is the total number of charged particles in the data sample and $<p_T>$ is the average \pT ~for a given
number of charged particles. 

For the multiplicity against pseudorapidity distribution, all MC model~\cite{atlas_mc} shapes seem to agree well with the data except for the normalization difference. However for the other two distributions, pre-LHC models show significant disagreement, especially in the low \pT ~and low multiplicity region, where one expects to have the largest contribution from diffractive events. The new ATLAS Minimum Bias Tune (AMBT)~\cite{ambt}, obtained by attempting to fit the charged particle multiplicity distributions in a diffraction limited phase-space, and also using the plateau of the underlying event distributions to be shown next, results in a significant improvement.

%%UE

\begin{figure}[htbp]
\centering
\subfigure[]{
\includegraphics[scale=0.25]{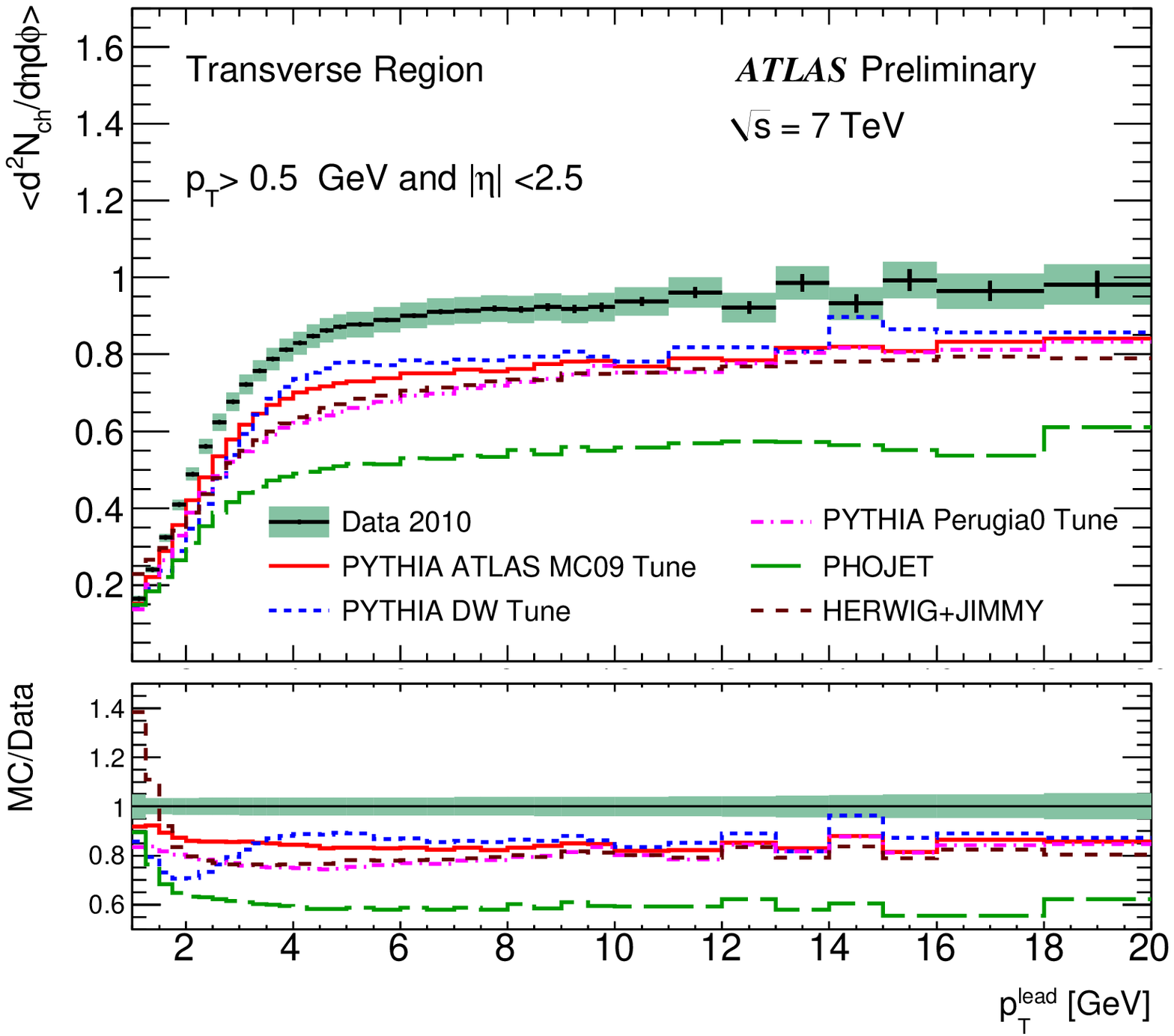}
%\label{fig:ue_pnchg}
}
\subfigure[]{
\includegraphics[scale=0.25]{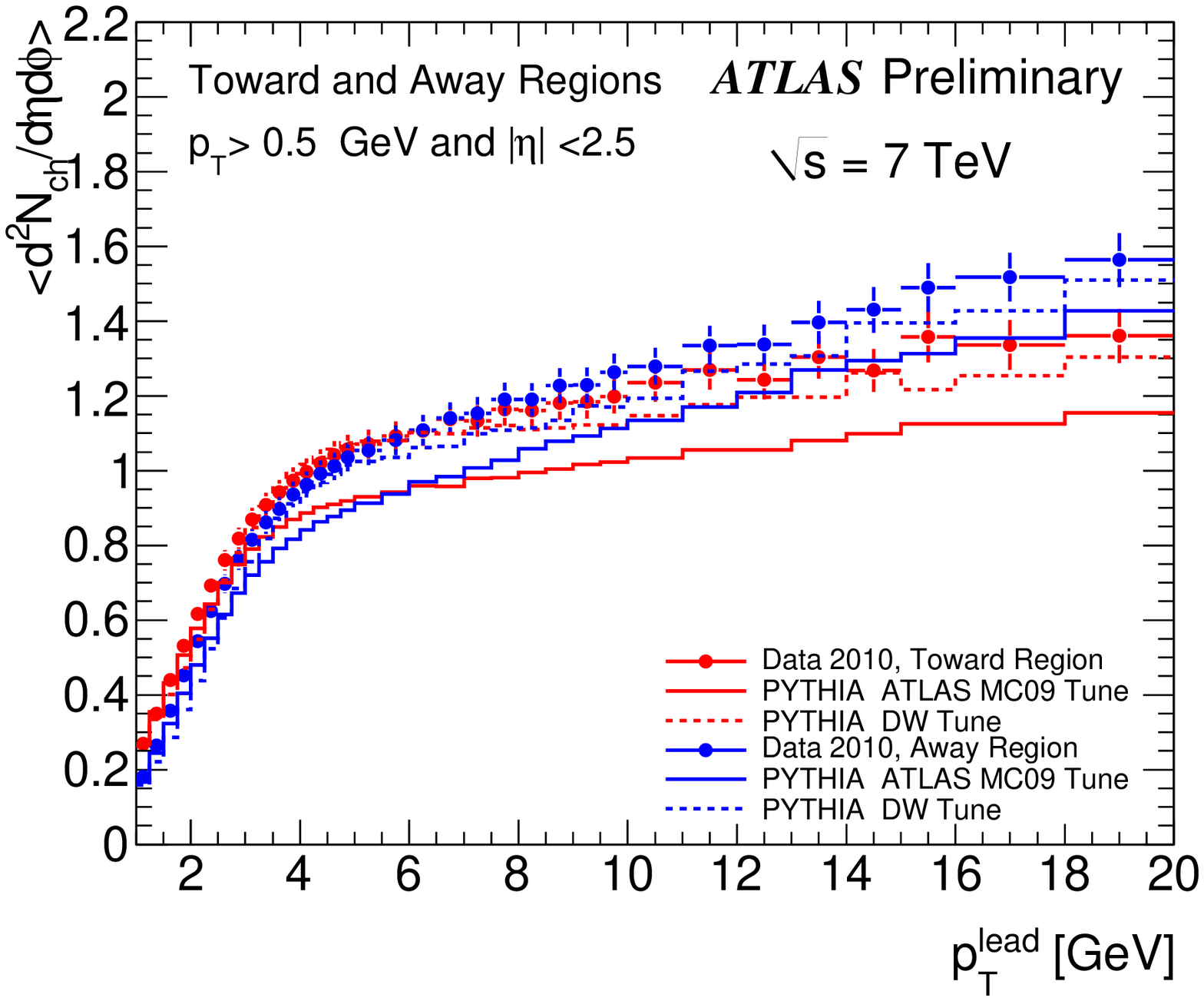}
%\label{fig:ue_lol}
}
\subfigure[]{
\includegraphics[scale=0.25]{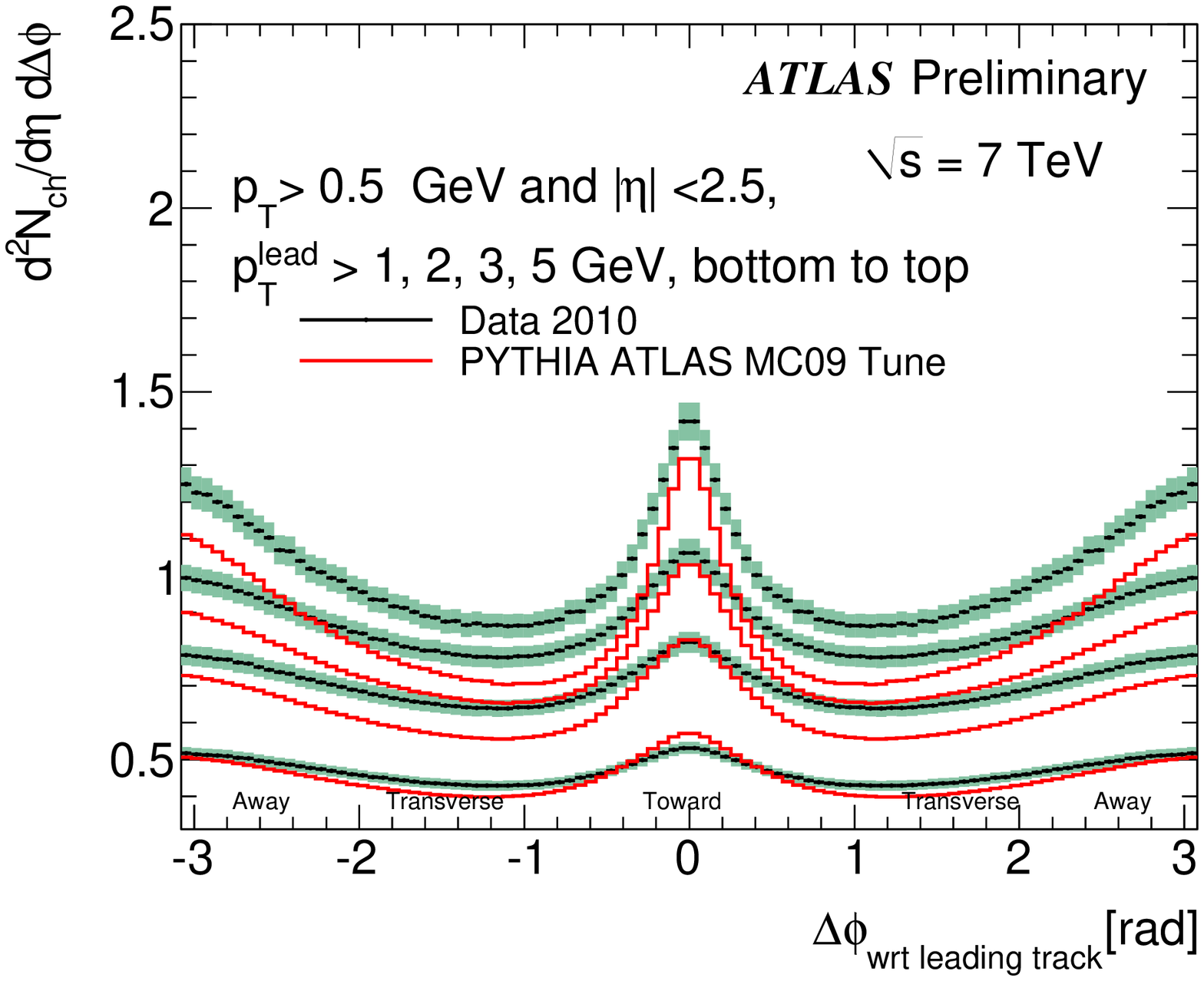}
%\label{fig:ue_deltaphi}
}
\caption[]{The the density of the charged particles as a function of the leading track \pT is shown in the transverse region in (a) and in toward and away regions in (b). The  $\phi$ distribution of track densities with respect to the leading track is shown in (c).}
\label{fig:ue}
\end{figure}

The underlying event is defined as everything except the hard scattered part, so it consists of the 
accompanying beam--beam remnants (BBR) and the multiple parton interactions (MPI)~\cite{ue_cdf}.
The underlying event also receives contributions from initial and final state parton QCD radiation, and on
an event-by-event basis it is impossible to separate them out. 
The direction of the leading track with \pT $> 1$ GeV, is used to isolate regions
$\eta-\phi$ space that are sensitive to the underlying event.
We define $|\Delta\phi| < 60^{\circ}$ as the `towards region', $60^{\circ} < |\Delta\phi| < 120^{\circ}$ as the `transverse region'; and
$|\Delta\phi| > 120^{\circ}$ as the `away region', where the azimuthal angle $\Delta\phi$ is relative to the leading track.
The transverse region is almost perpendicular to the plane of 2-2 hard scattering and therefore 
very sensitive to the underlying event. The data is corrected back to particle level by using a similar
correction procedure used in minimum bias analysis, and underlying event distributions~\cite{ue20} 
with \pT $> 0.5$ GeV and $|\eta| < 2.5$  are 
compared with predictions from different MC models in \FigRef{fig:ue}.

One of the observables sensitive to underlying event is the density of charged particles as a function of the leading track \pT. 
All the pre-LHC MC tunes considered show lower activity than the data in the transverse region.
The toward and away regions are dominated by jet-like activity, yielding gradually rising number densities.
In contrast, the number density in the transverse region appears to be independent of the energy scale defined by the \pT of the leading track once it reaches the plateau.   
It is observed that the charged particle density in the underlying event, in the plateau region 
is about a factor of two larger than the number of charged particles per unit rapidity seen in the
inclusive minimum-bias spectrum. Given that there is one hard
scattering it is more probable to have MPI, and hence, the underlying event has
more activity than minimum-bias. 
The angular distribution of the charged particle number with respect to the leading track for the event 
is also shown for different leading track \pT ~slices. 
These plots are symmetrized by reflecting then about $\phi=0$, and the leading track is excluded.
These distributions show a significant difference in shape between data and MC predictions.

%%% Delta Phi

\begin{figure}[htbp]
\centering
\subfigure[]{
\includegraphics[scale=0.2]{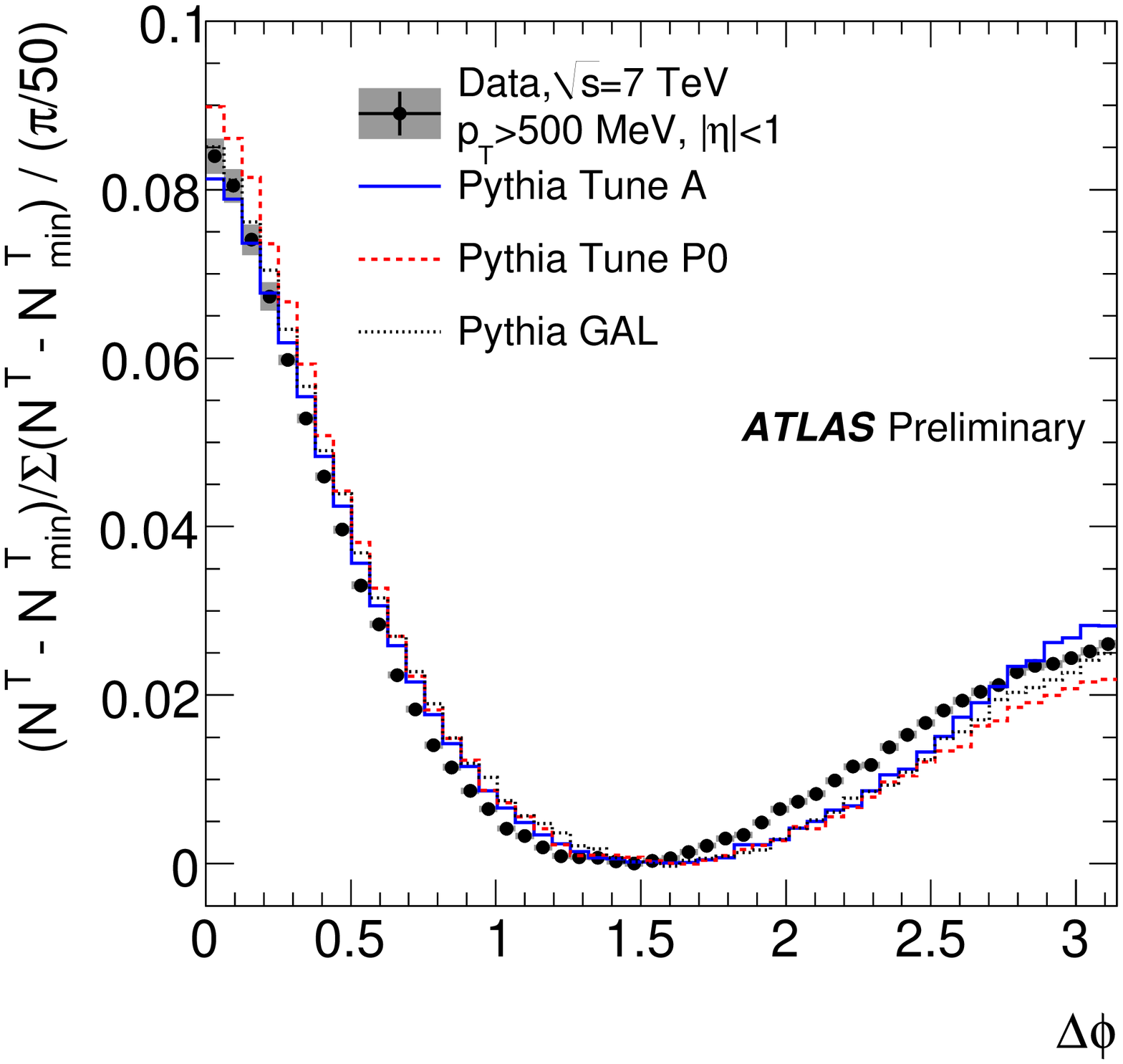}
%\label{fig:dphi1}
}
\subfigure[]{
\includegraphics[scale=0.2]{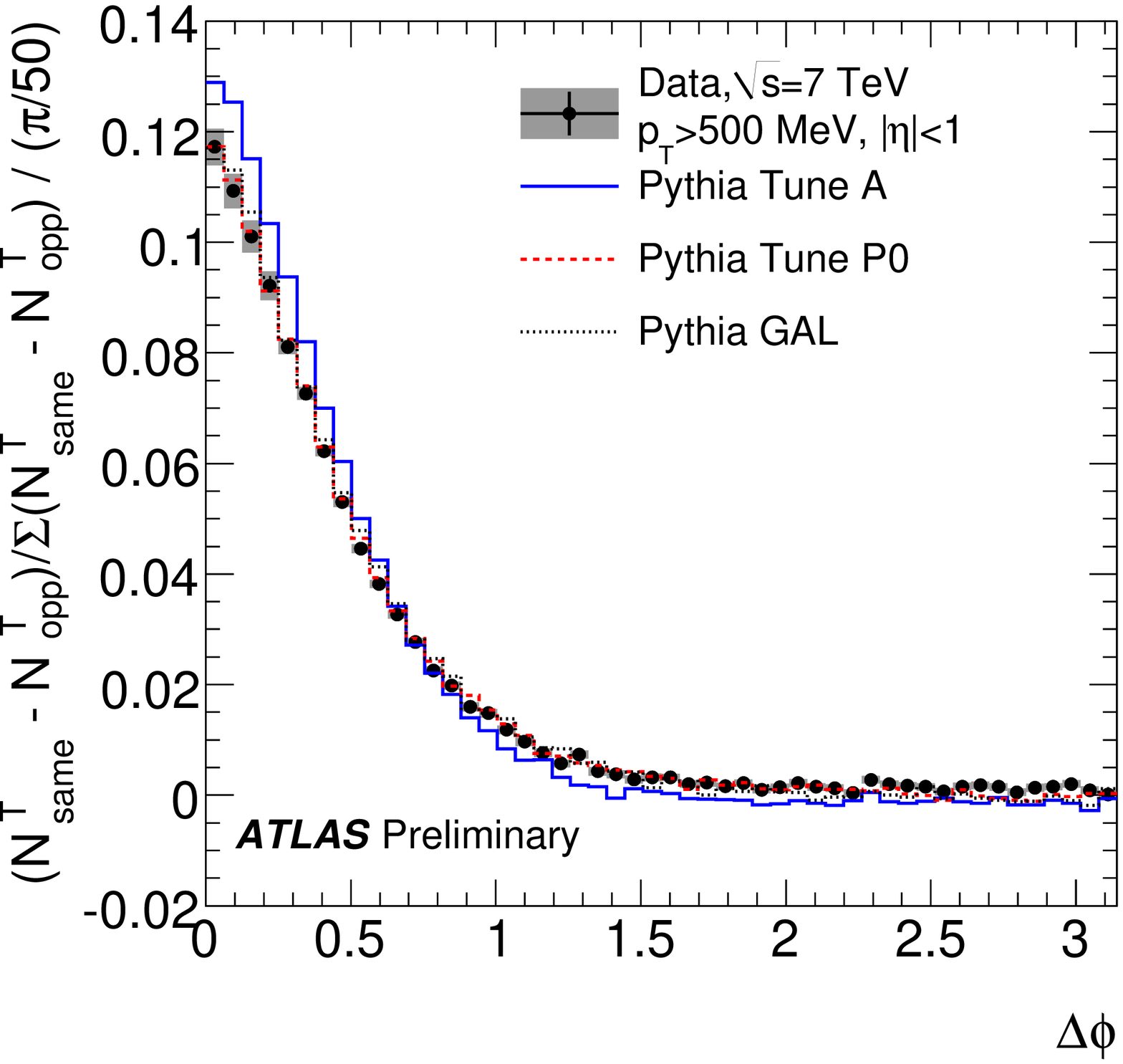}
%\label{fig:dphi2}
}
\subfigure[]{
\includegraphics[scale=0.2]{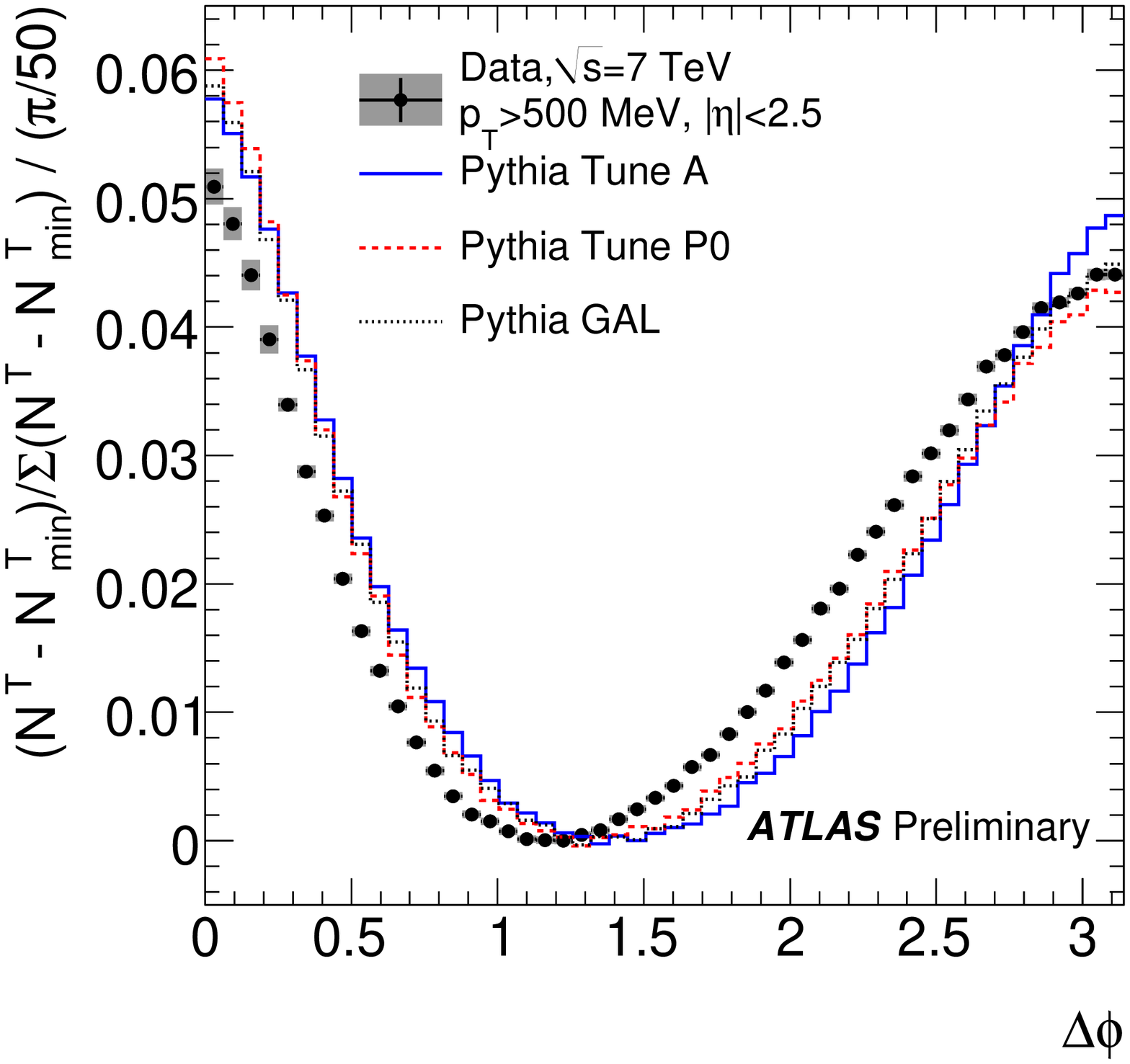}
%\label{fig:dphi3}
}
\subfigure[]{
\includegraphics[scale=0.2]{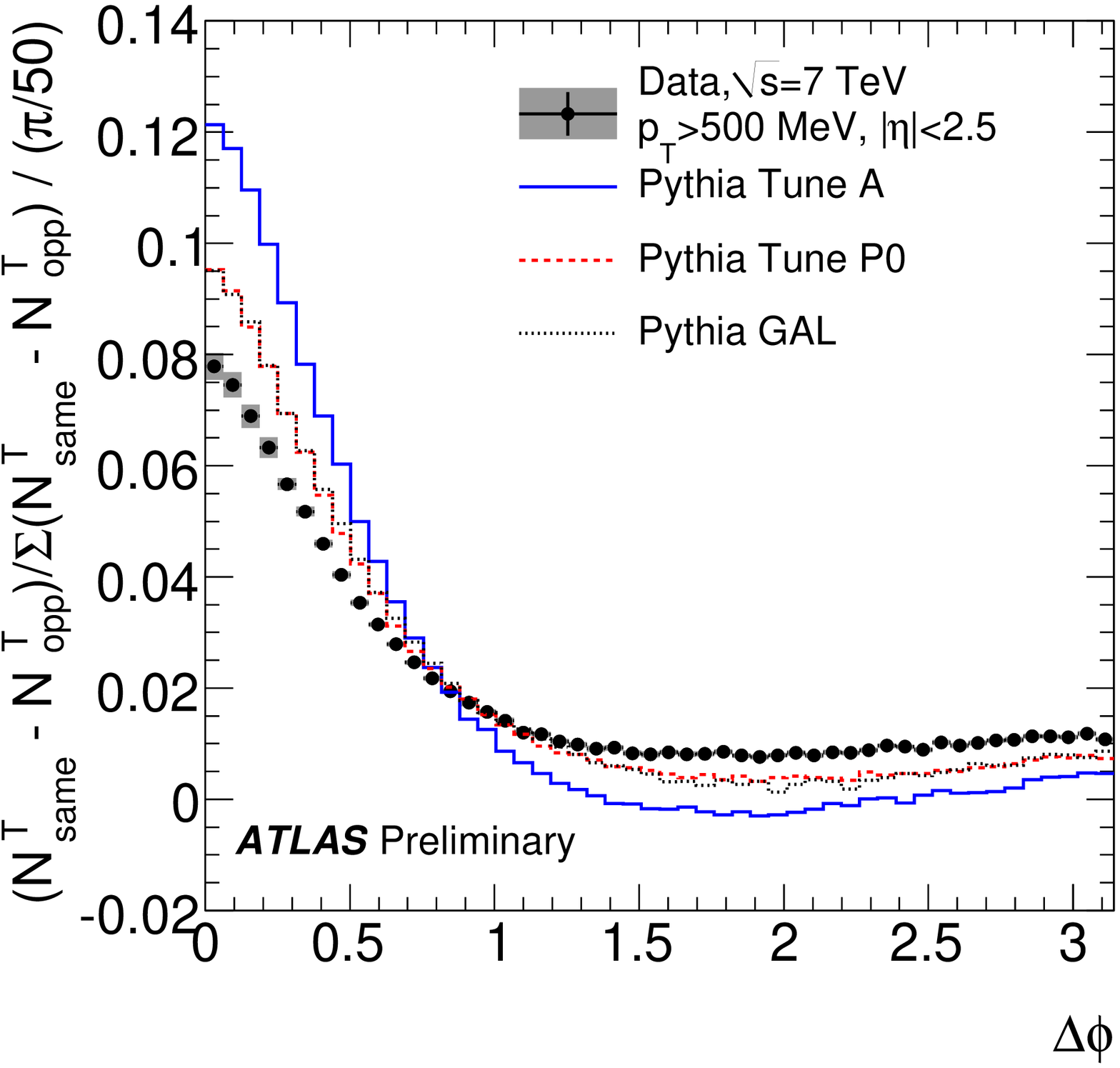}
%\label{fig:dphi4}
}
\caption[]{The $\Delta \phi$ crest shape obtained by subtracting the minimum are shown in (a) and (c), while those obtained by the subtracting the `opposite from same' are shown in (b) and (d). The left two plots are for $|\eta| < 1.0$ and the right two plots are for $|\eta| < 2.5$.}
\label{fig:dphi}
\end{figure}

A complementary way~\cite{dphi} to look at the angular correlation is by either subtracting the minimum of the distribution (determined by a second-order polynomial fit), or by subtracting the opposite side distribution (defined according to if pseudorapidity has the same or the opposite sign as the leading track) from the same side distribution and normalizing to unity. In \FigRef{fig:dphi}, it is seen that the models are better at lower $\eta$ than at higher.

\section{JET PHYSICS MEASUREMENTS}

Jets are distinctive signature of short-distance (`hard') interactions between partons, 
and probes different aspects of high \pT ~physics. In ATLAS jets from both charged particle tracks and 
calorimeters clusters at $\sqrt{s} =$ 7 TeV using anti-$k_T$~\cite{antikt} algorithm with jet size of 0.4 and 0.6 are looked at. 
Calorimeter jets are chosen with at least one jet within the kinematic region $p_T > 60$ GeV and $|y| < 2.8$ (other jets with $p_T> 30$ GeV).
They are corrected back to particle level, and jet energy scale corrections based on MC are applied to the measured jets in EM scale.
Jet energy scale uncertainty is smaller than 7\% for $p_{T,jet} > 100$ GeV.
All results are shown here do not include the 11\% systematic uncertainty on the luminosity measurement.

%%Track Jet

\begin{figure}[htbp]
\centering
\subfigure[]{
\includegraphics[scale=0.25]{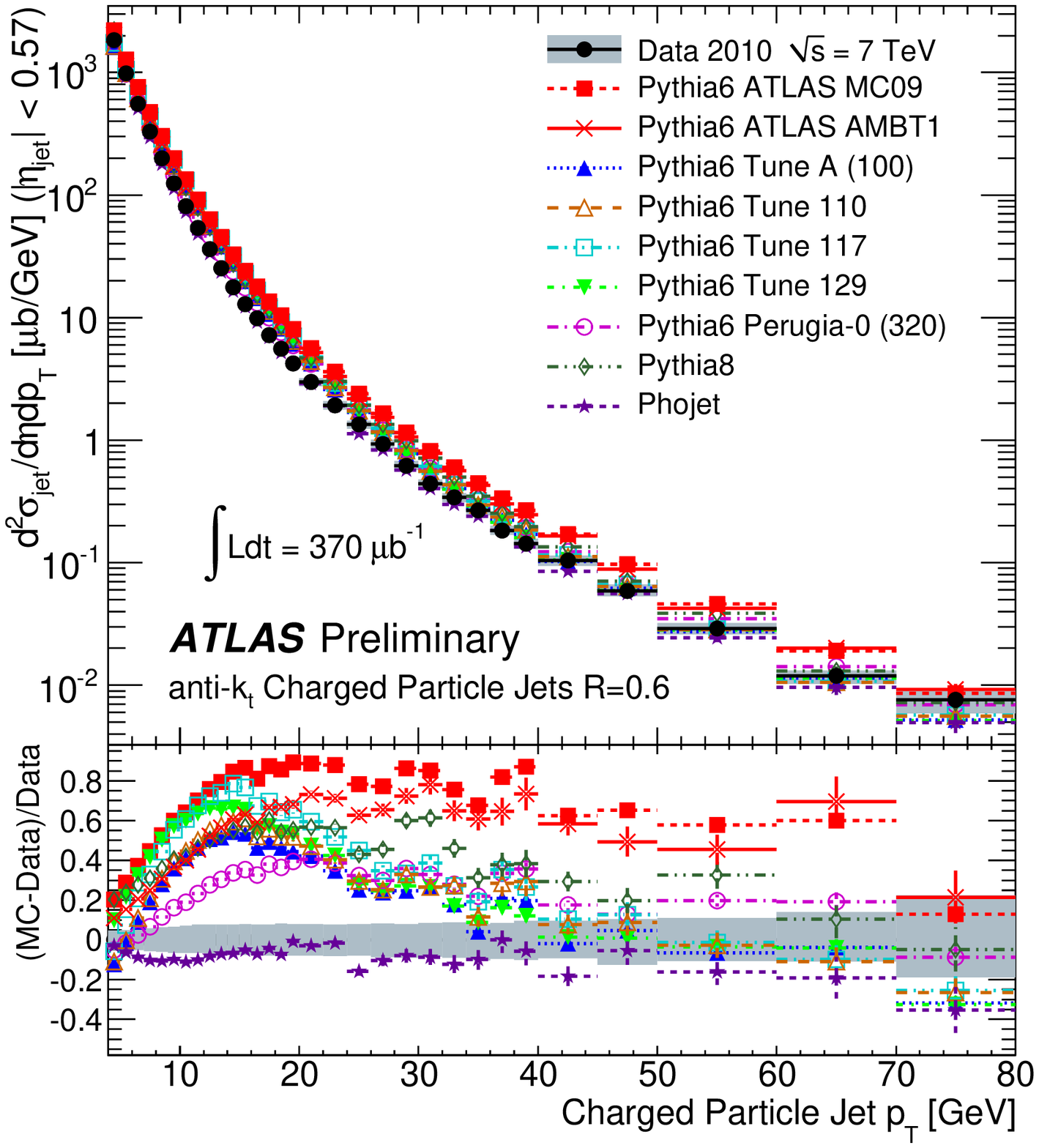}
%\label{fig:tj_cs}
}
\subfigure[]{
\includegraphics[scale=0.25]{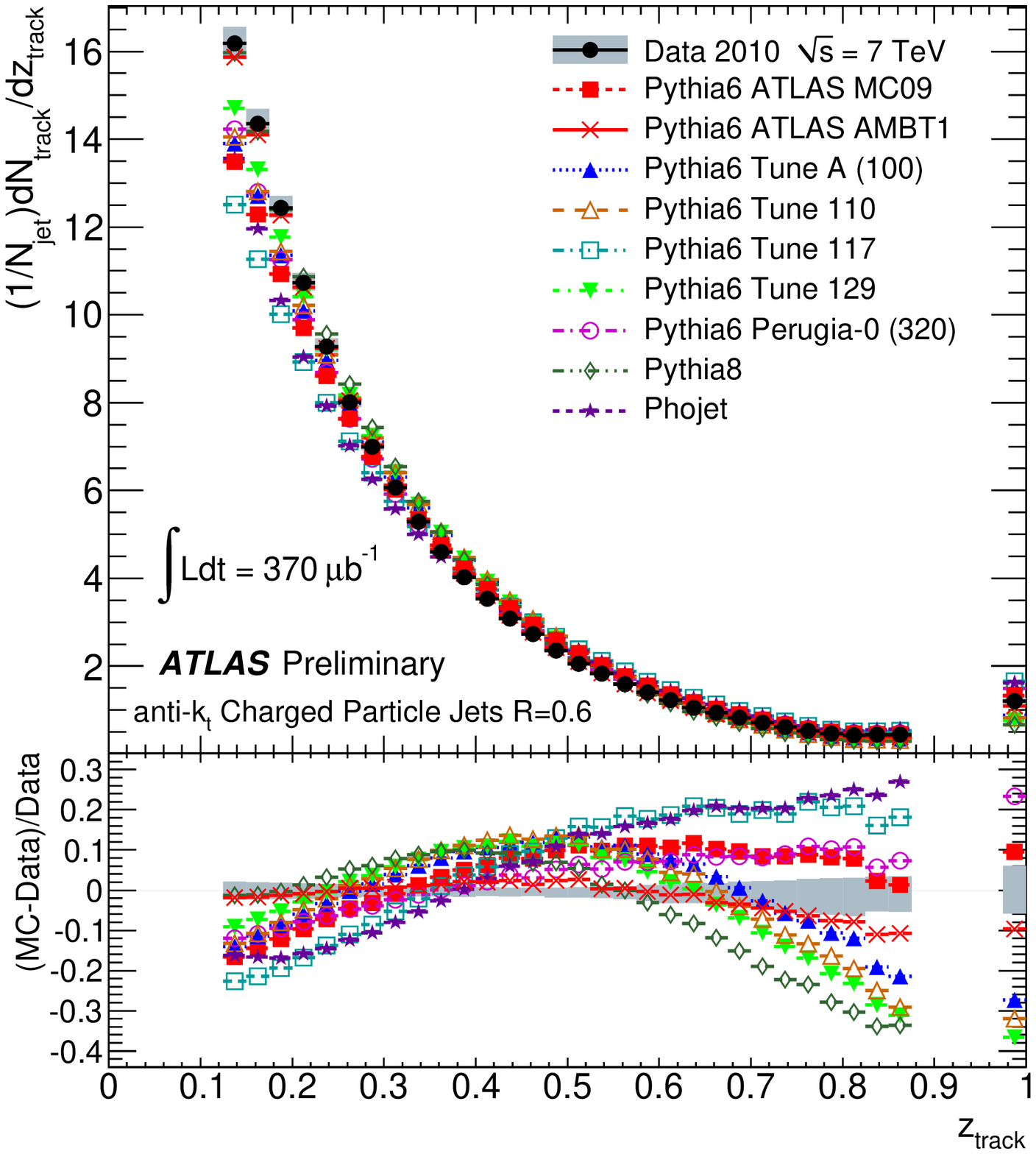}
%\label{fig:tj_fr1}
}
\subfigure[]{
\includegraphics[scale=0.25]{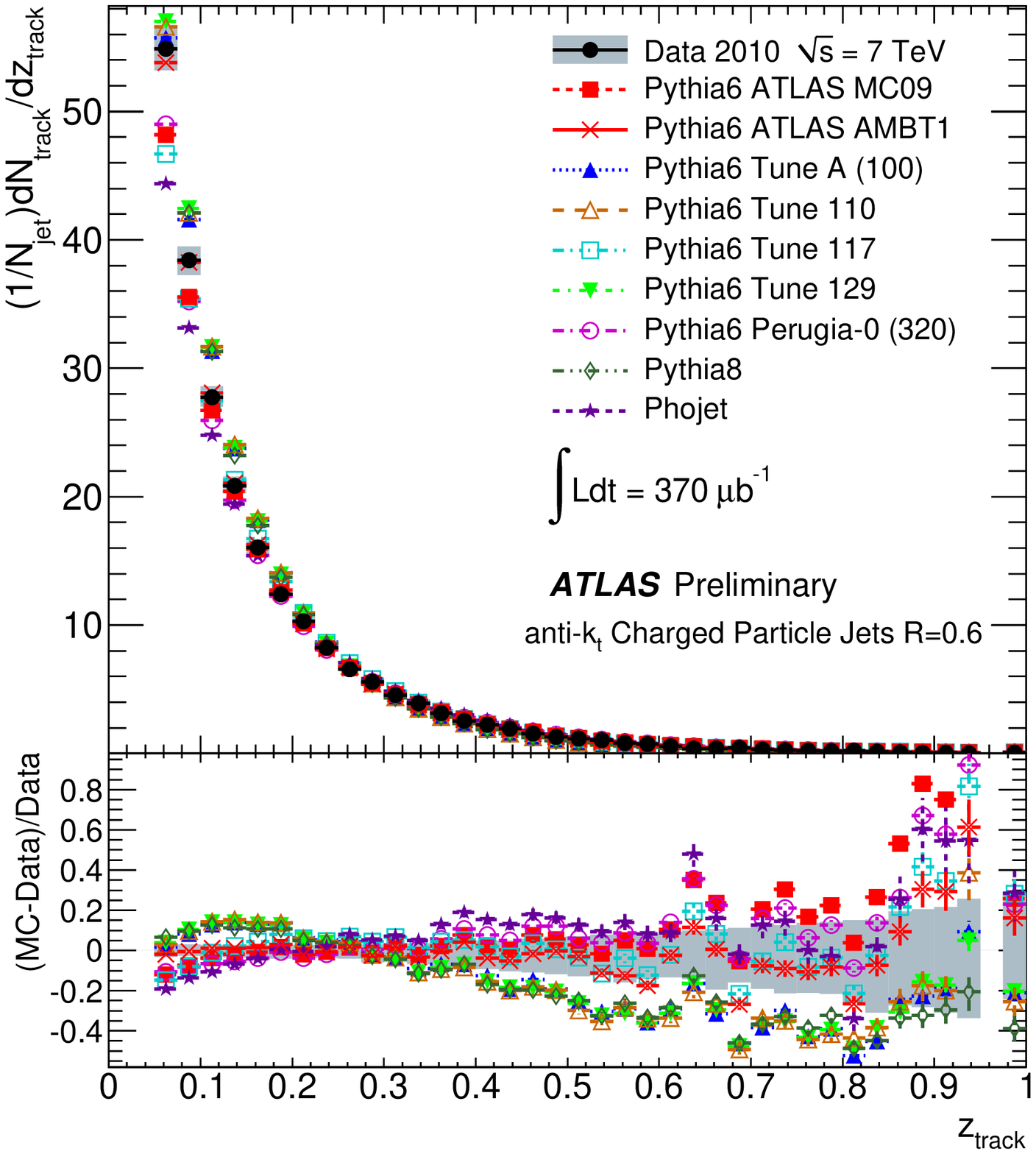}
%\label{fig:tj_fr2}
}
\caption[]{Corrected track-jet differential cross section is shown in (a). Corrected fragmentation function is shown in (b) for jet \pT from 4 GeV to 6 GeV and in (c) for jet \pT from 15 GeV to 24 GeV.}
\label{fig:tj}
\end{figure}

Trackjets are constructed with tracks having a minimum \pT ~of 500 MeV, and jets have \pT $>$ 4 GeV and $|\eta| <$ 0.57~\cite{tj}. The momentum distribution of tracks inside a jet is given by the fragmentation function $f(z)$ defined as the probability that a particle carries fraction $z$ of the jet momentum.
In \FigRef{fig:tj} corrected jet cross section and fragmentation function in different regions of jet \pT are shown for anti-$k_T$ jets with $R=0.6$.
The difference in crosssection between data and MC is mostly due to normalization. No models describes the fragmentation function data well for all jet momenta and all values of $z$. These distributions can be used improve phenomenological models of jet production and fragmentation that are implemented in MC generators.

%%%Inclusive jet

\begin{figure}[htbp]
\centering
\subfigure[]{
\includegraphics[scale=0.205]{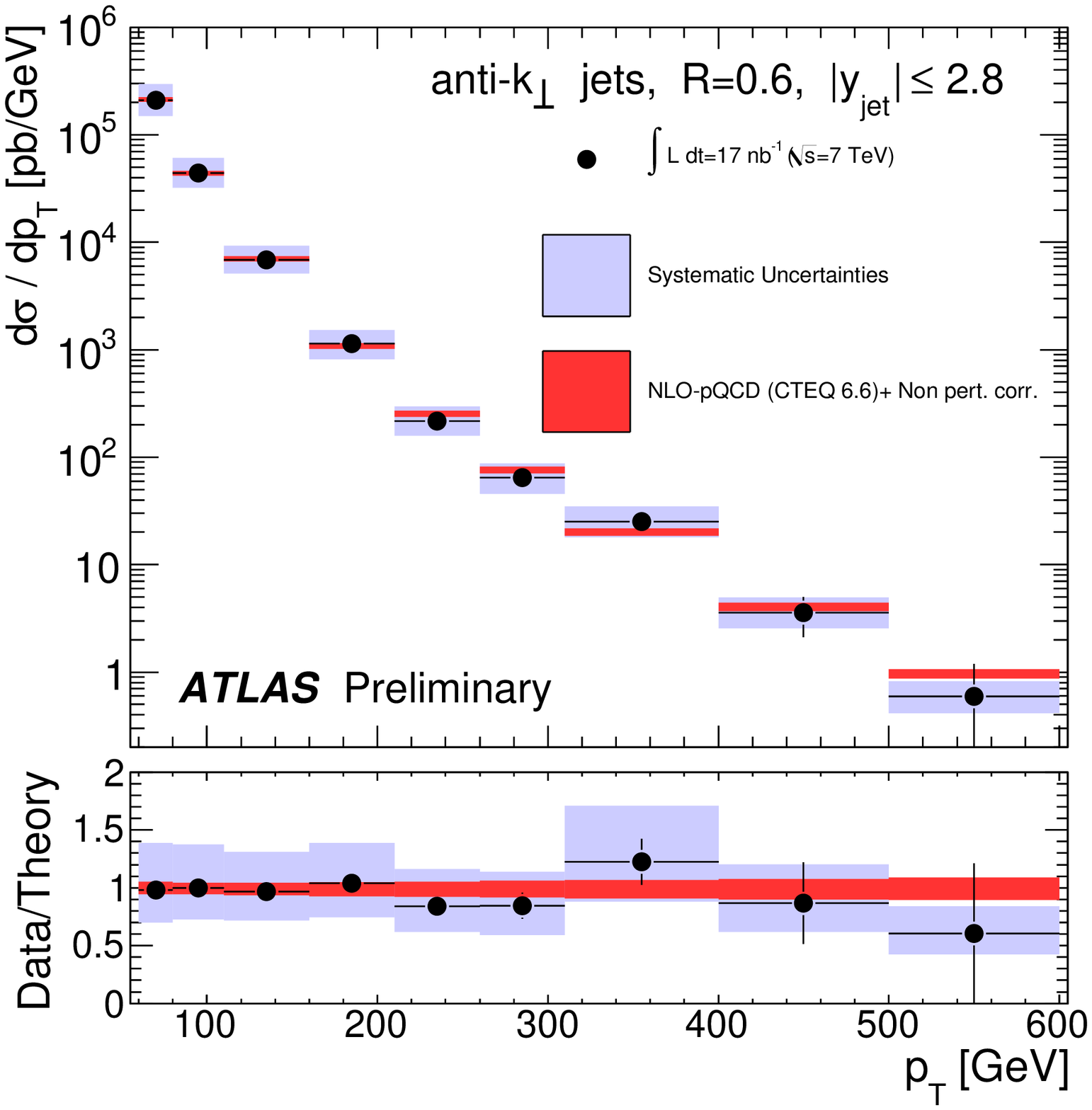}
%\label{fig:ij_cs1}
}
\subfigure[]{
\includegraphics[scale=0.205]{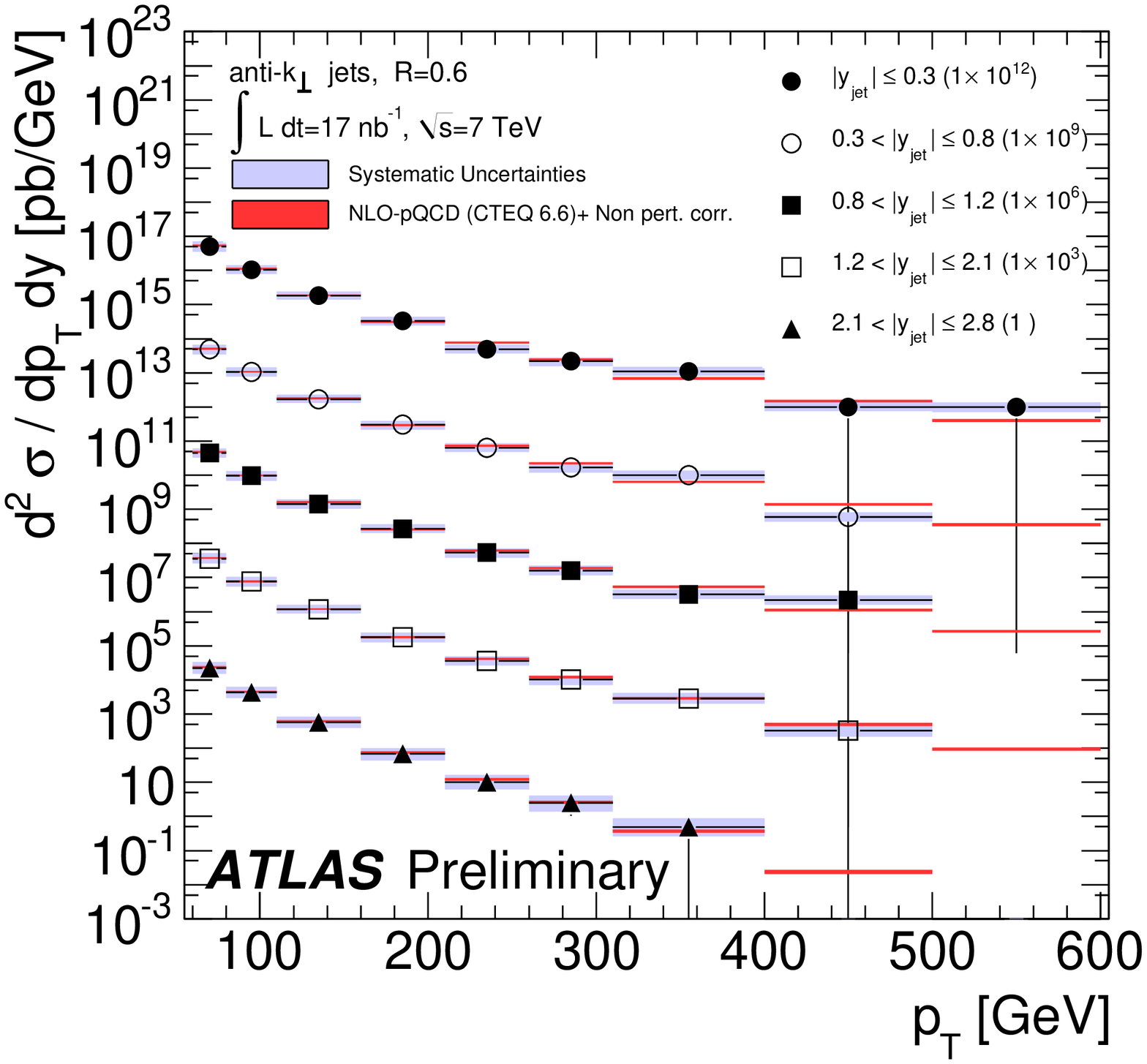}
%\label{fig:ij_cs2}
}
\subfigure[]{
\includegraphics[scale=0.205]{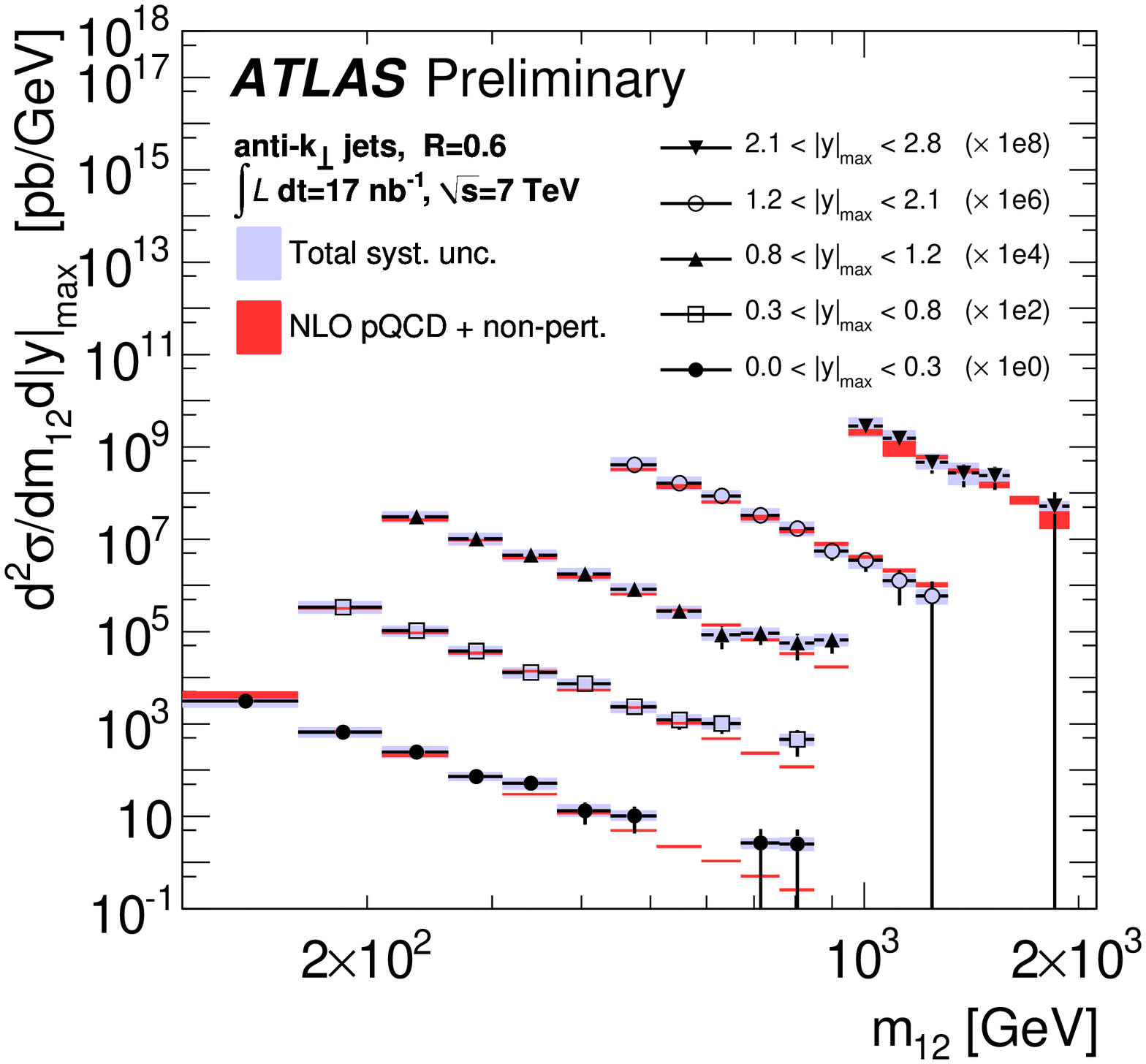}
%\label{fig:dj_cs1}
}
\subfigure[]{
\includegraphics[scale=0.205]{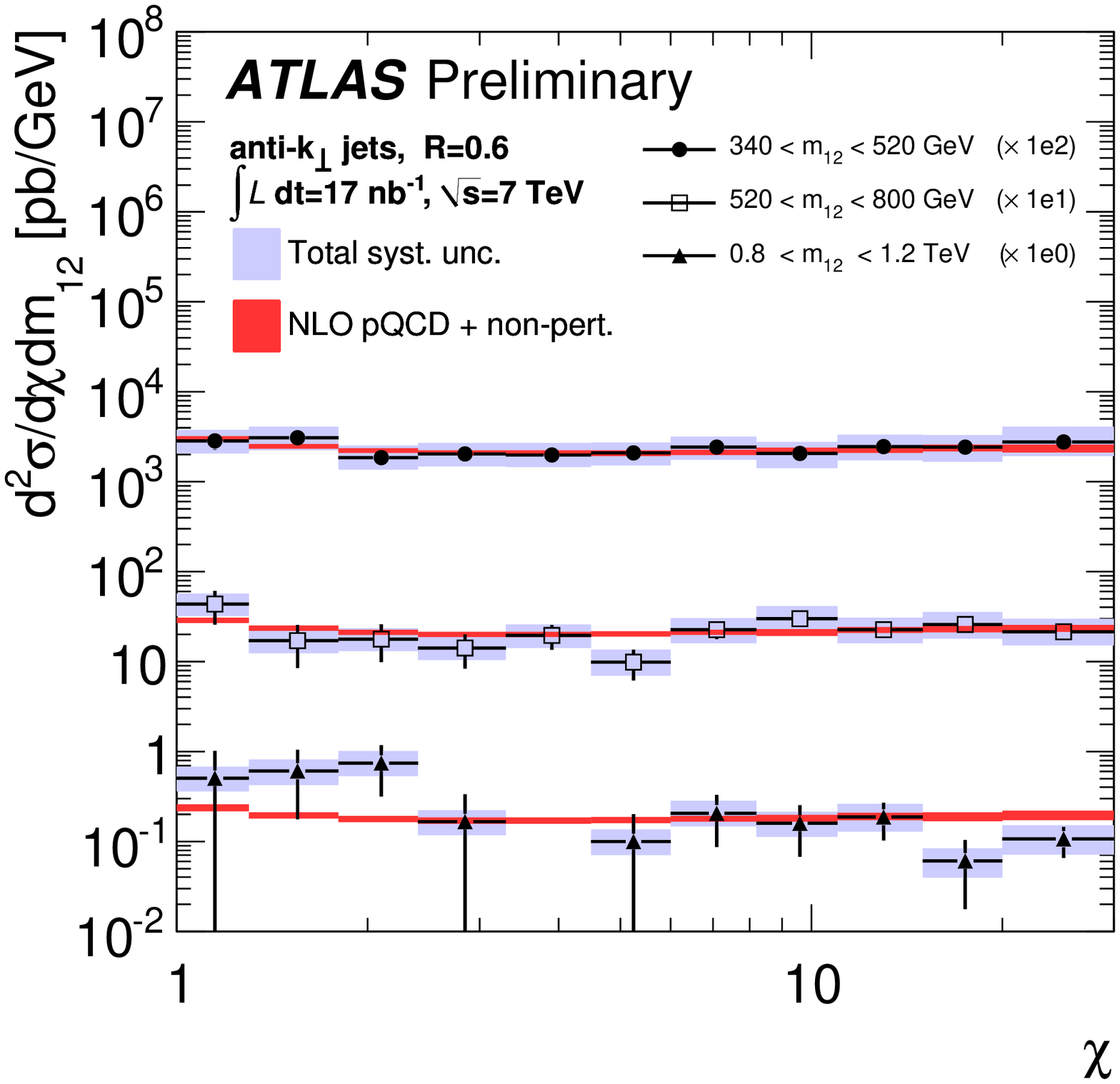}
%\label{fig:dj_cs2}
}
\caption[]{Inclusive jet differential cross-section as a function of jet \pT integrated over the full rapidity region is shown in (a) and inclusive jet double-differential cross-section as a function of jet \pT in different regions of |y| is shown in (b). In (c), dijet double-differential cross section as a function of dijet mass is shown and in (d), dijet double-differential cross section as a function of angular variable $\chi$ in different bins of dijet mass is shown.}
\label{fig:ij}
\end{figure}

Inclusive and dijet cross-sections~\cite{inj} using the anti-$k_T$ algorithm with $R=0.6$ are shown in \FigRef{fig:ij}. Inclusive jet differential cross section as a function of jet \pT ~integrated over the full region $|y|<2.8$ represents the probability to observe a jet as a function of the jet \pT. single-jet cross-section is less sensitive to final state topology. The data are compared to NLO QCD calculations to which soft QCD corrections have been applied. The theory uncertainty is the quadratic sum of uncertainties from the choice of renormalization and factorization scales, parton distribution functions, $\alpha_s$, and the modeling of soft QCD effects. The cross section falls by more than four orders of magnitude over the range. Data are theory are consistent for the full region, as well as for the inclusive jet double-differential cross section in different regions of  rapidity. Dijet double-differential cross section as a function of dijet mass, binned in the maximum rapidity of the two leading jets, $|y|_max = max(|y_1|, |y_2|)$, reaches masses up to $\sim$ 2 TeV, overtaking Tevatron analysis in mass reach. This is a probe for exotic resonances, that tend to decay to high \pT ~central jets. QCD produces more low \pT ~forward jets separated by large opening angle. Good agreement between data and NLO pQCD prediction is observed. Dijet double-differential cross section is also measured as a function of angular variable $\chi$, which measures scattering angle in dijet center-of-mass frame. Data and theory is seen to be consistent in all mass regions. This distribution is designed to look for new physics which appears as a shape difference, namely an enhancement at small scattering angles, arising from contact interactions e.g. compositeness or gravitational scattering, whereas QCD is roughly flat as we see. This is one most sensitive observable to new physics searches that can be compared to a theory prediction.

%%%Multijet

\begin{figure}[htbp]
\centering
\subfigure[]{
\includegraphics[scale=0.24]{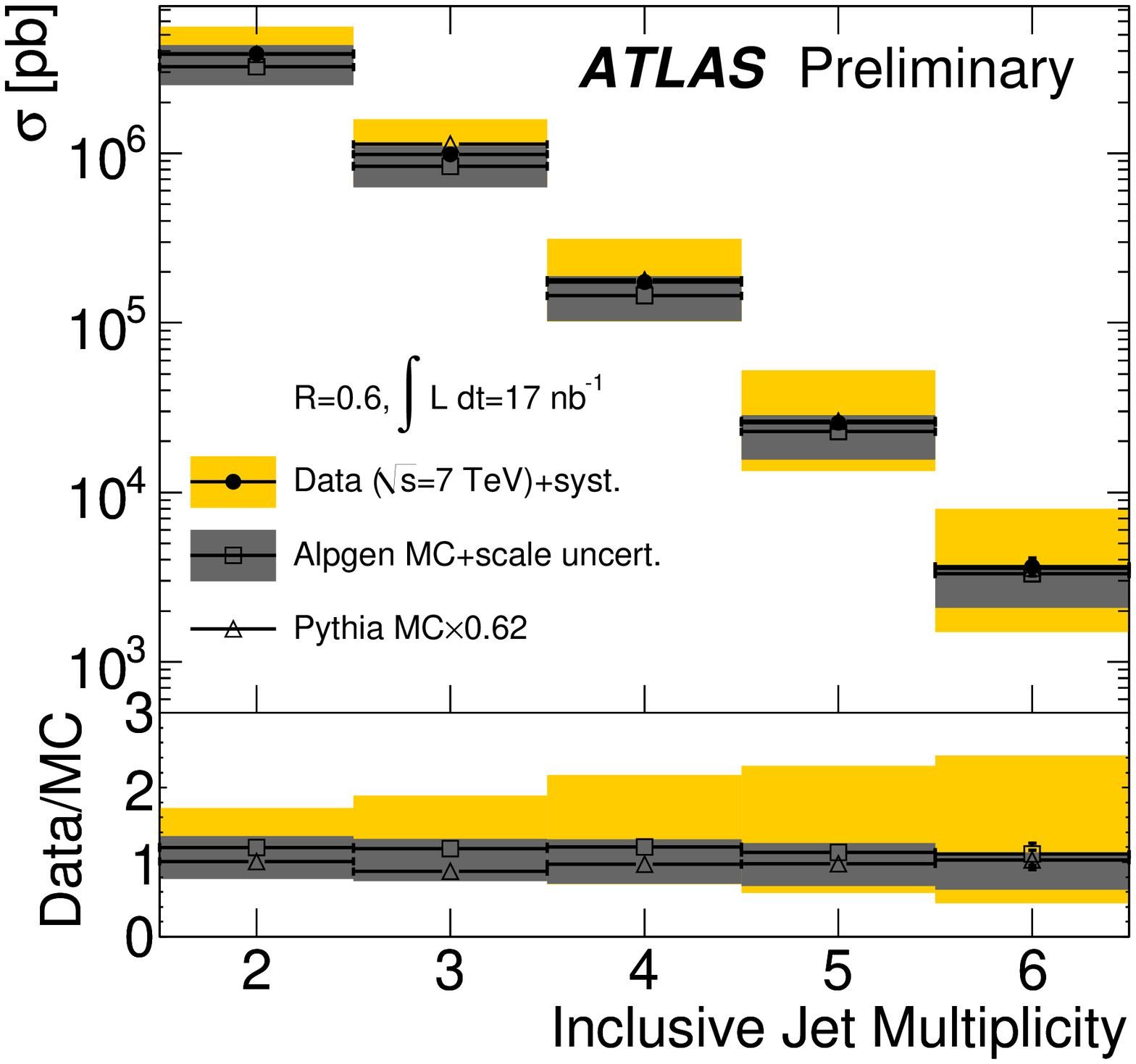}
%\label{fig:mj_cs1}
}
\subfigure[]{
\includegraphics[scale=0.24]{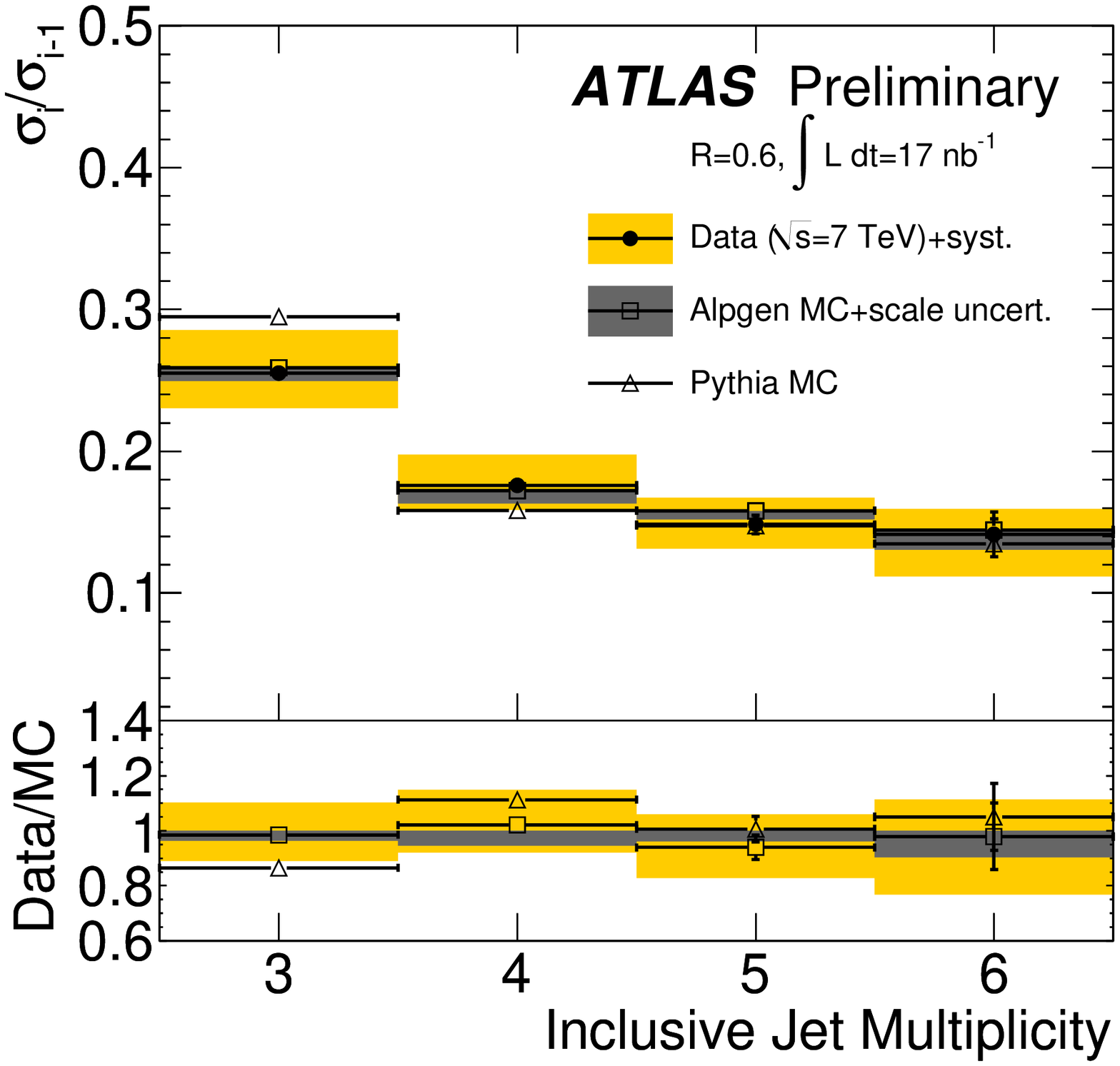}
%\label{fig:mj_cs2}
}
\subfigure[]{
\includegraphics[scale=0.24]{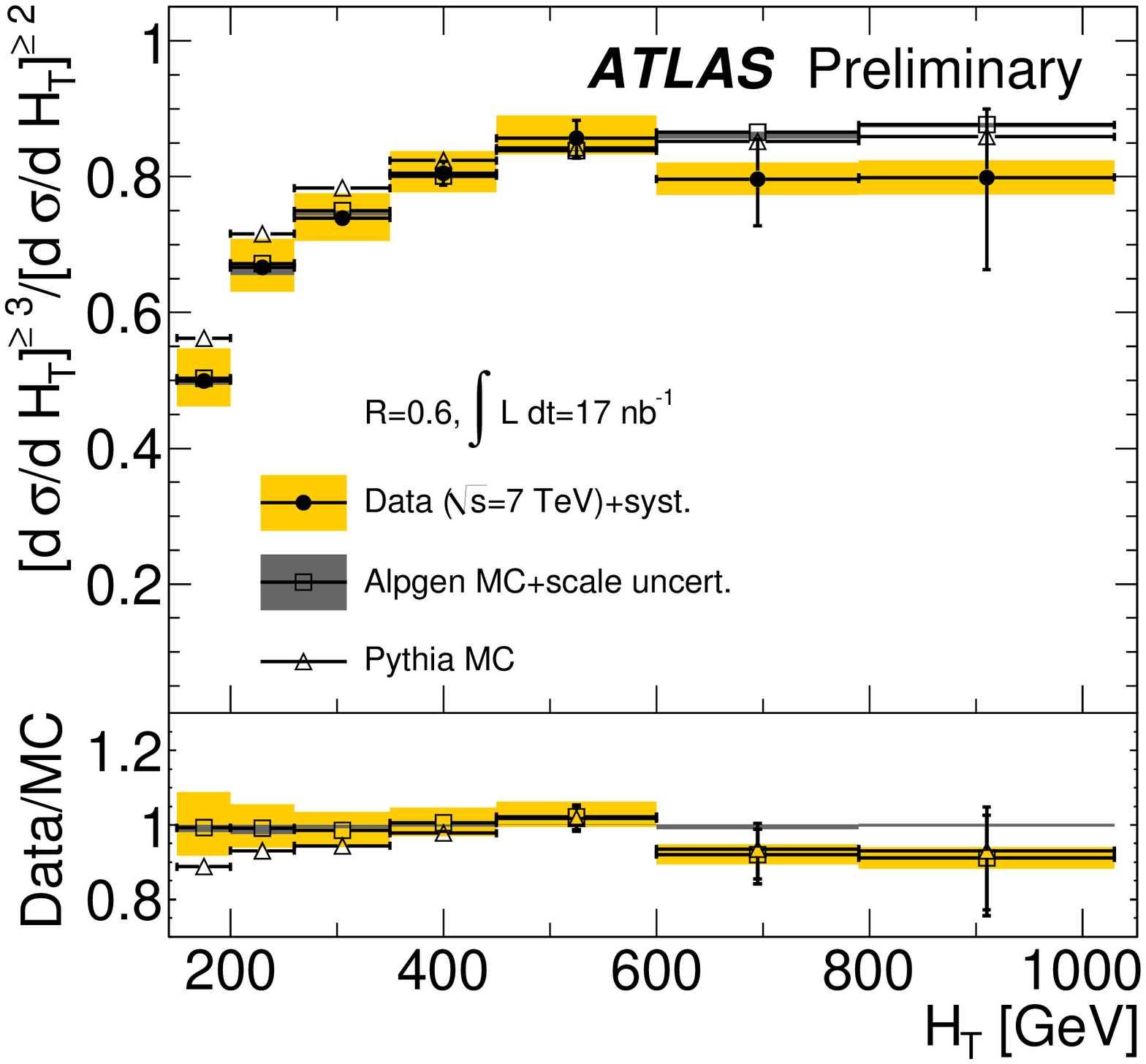}
%\label{fig:mj_cs3}
}
\caption[]{Inclusive jet multiplicity spectrum corrected to particle level as measured in data is shown in (a). Ratio of the n-jet cross section to the (n-1)-jet cross section for n varying from 3 to 6 as measured in data is shown in (b). In (c) the ratio of the 3-jet cross section to the 2-jet cross section as a function of $H_T$ as measured in data is shown.}
\label{fig:mj}
\end{figure}

Inclusive jet multiplicity and N-jet to N-1-jet cross-section ratios as a function of number of jets~\cite{mj} is shown in \FigRef{fig:mj}. Also the ratio of the 3-jet cross section to the 2-jet cross section as a function of the event level variable $H_T =\Sigma_{jets} p_T$, is shown.
The PYTHIA~\cite{pythia} simulation has been normalized to data, using the first bin to compare the shapes. The range spanned by the ALPGEN~\cite{alpgen} prediction when varying the renormalization and factorization scales simultaneously is shown as the error band. Data agree with both MC generators within the systematic uncertainties. Cross-section ratios reduce systematic uncertainties significantly and good agreement is seen with ALPGEN, but there is some disagreement with PYTHIA at low $H_T$. Ratio of 3-jet and 2-jet cross-sections in $H_T$ is a direct probe of $\alpha_s$.

%%%Gap Fraction

\begin{figure}[htbp]
\centering
\subfigure[]{
\includegraphics[scale=0.205]{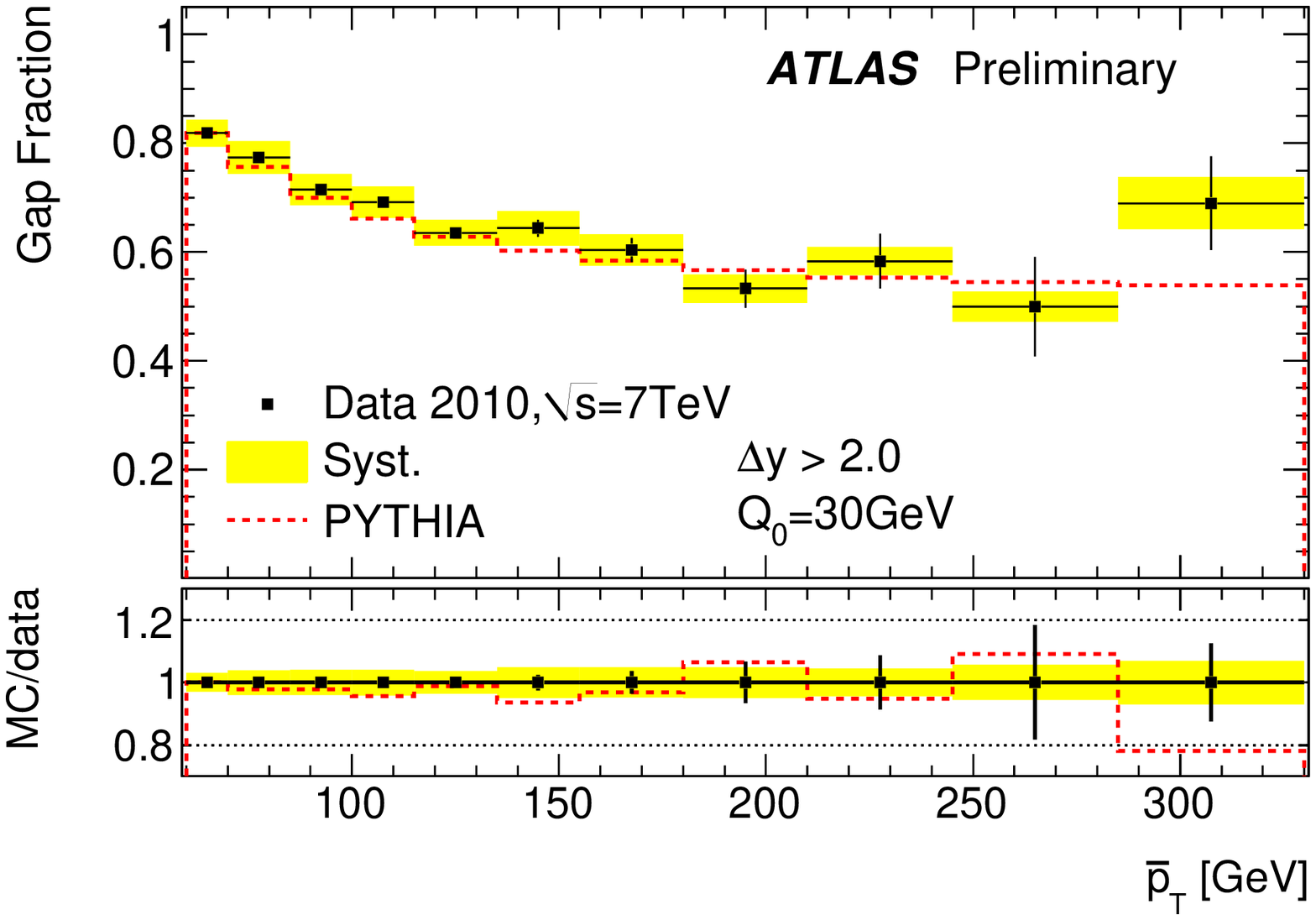}
%\label{fig:gap1}
}
\subfigure[]{
\includegraphics[scale=0.205]{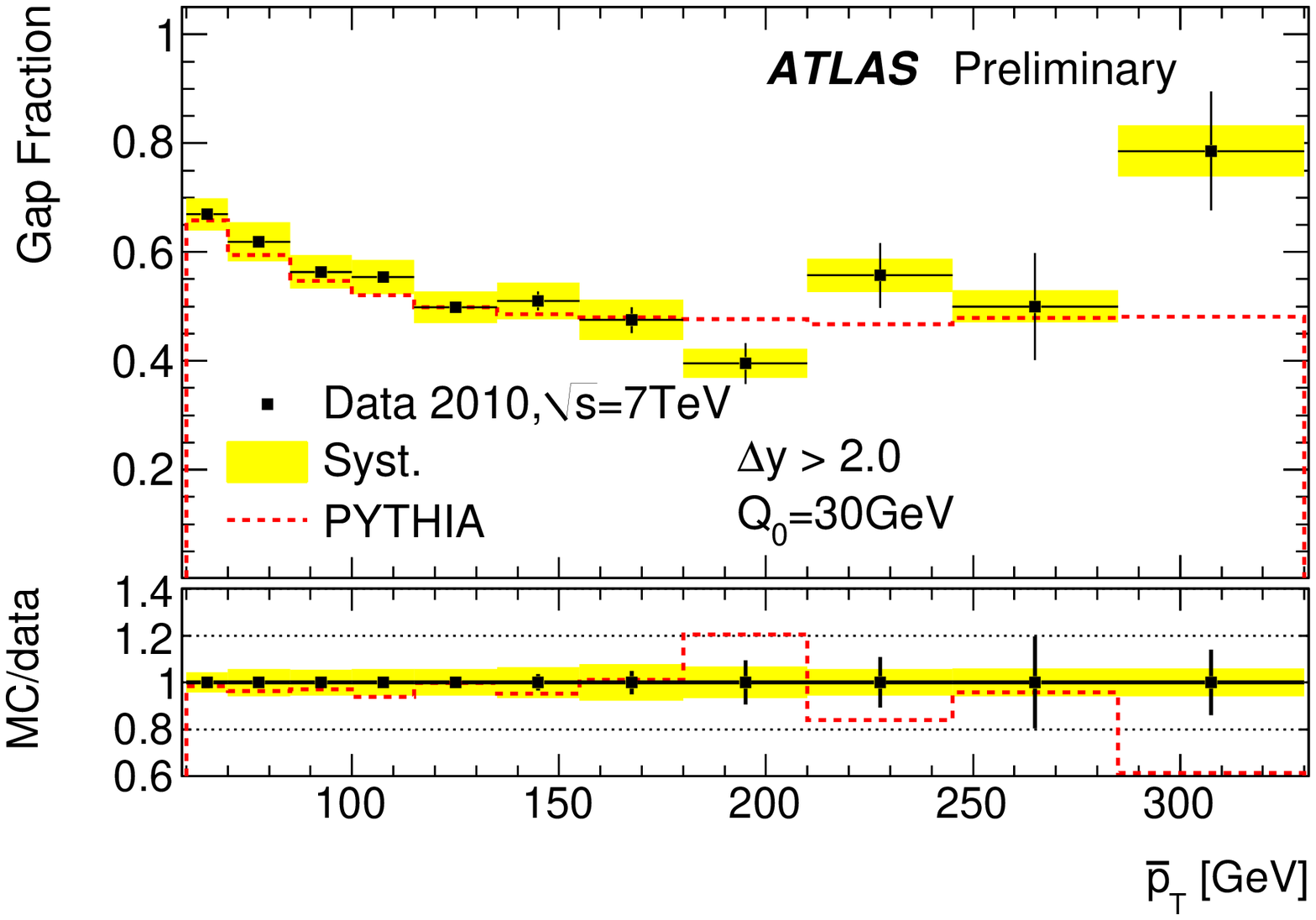}
%\label{fig:gap2}
}
\subfigure[]{
\includegraphics[scale=0.205]{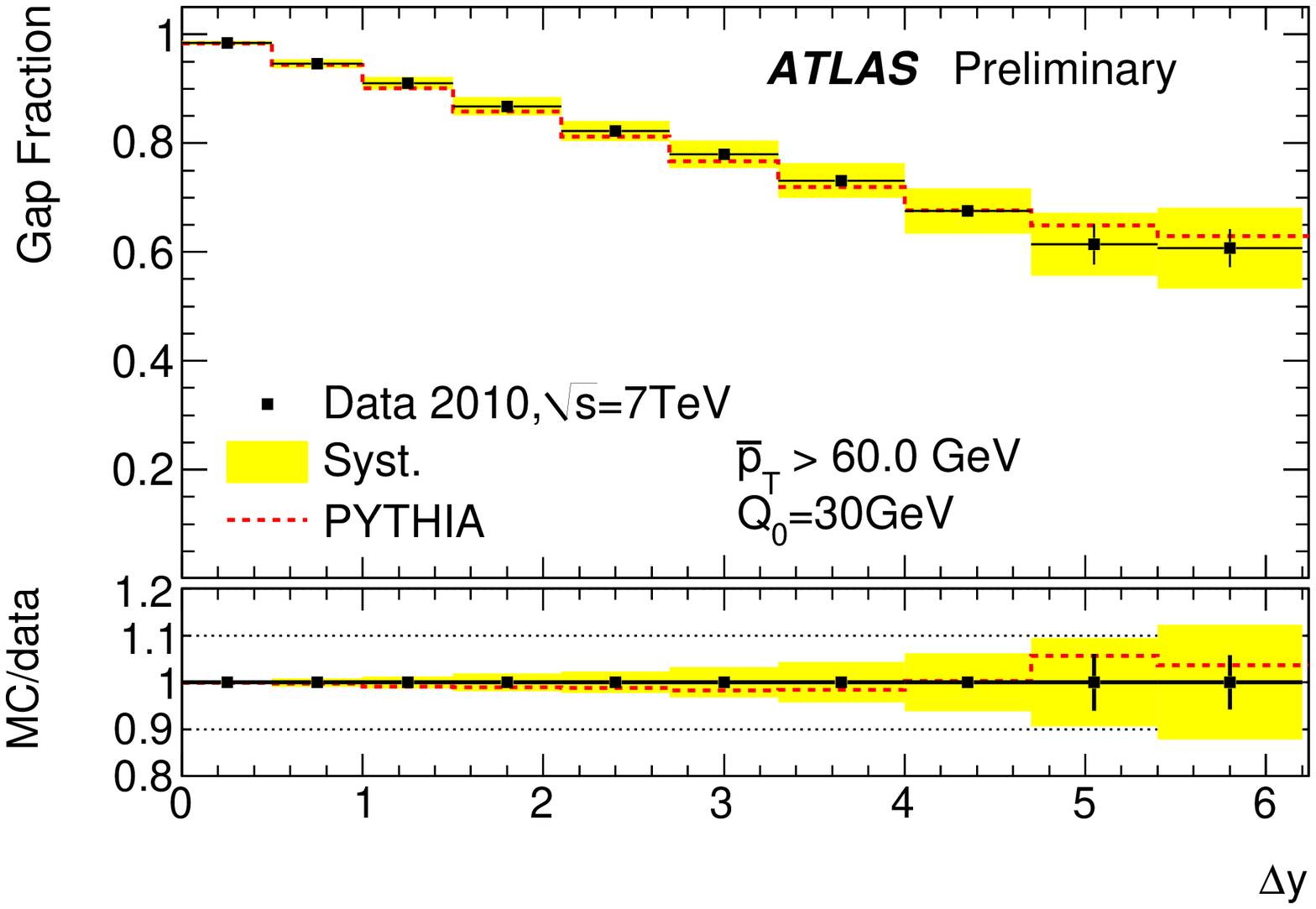}
%\label{fig:gap3}
}
\subfigure[]{
\includegraphics[scale=0.205]{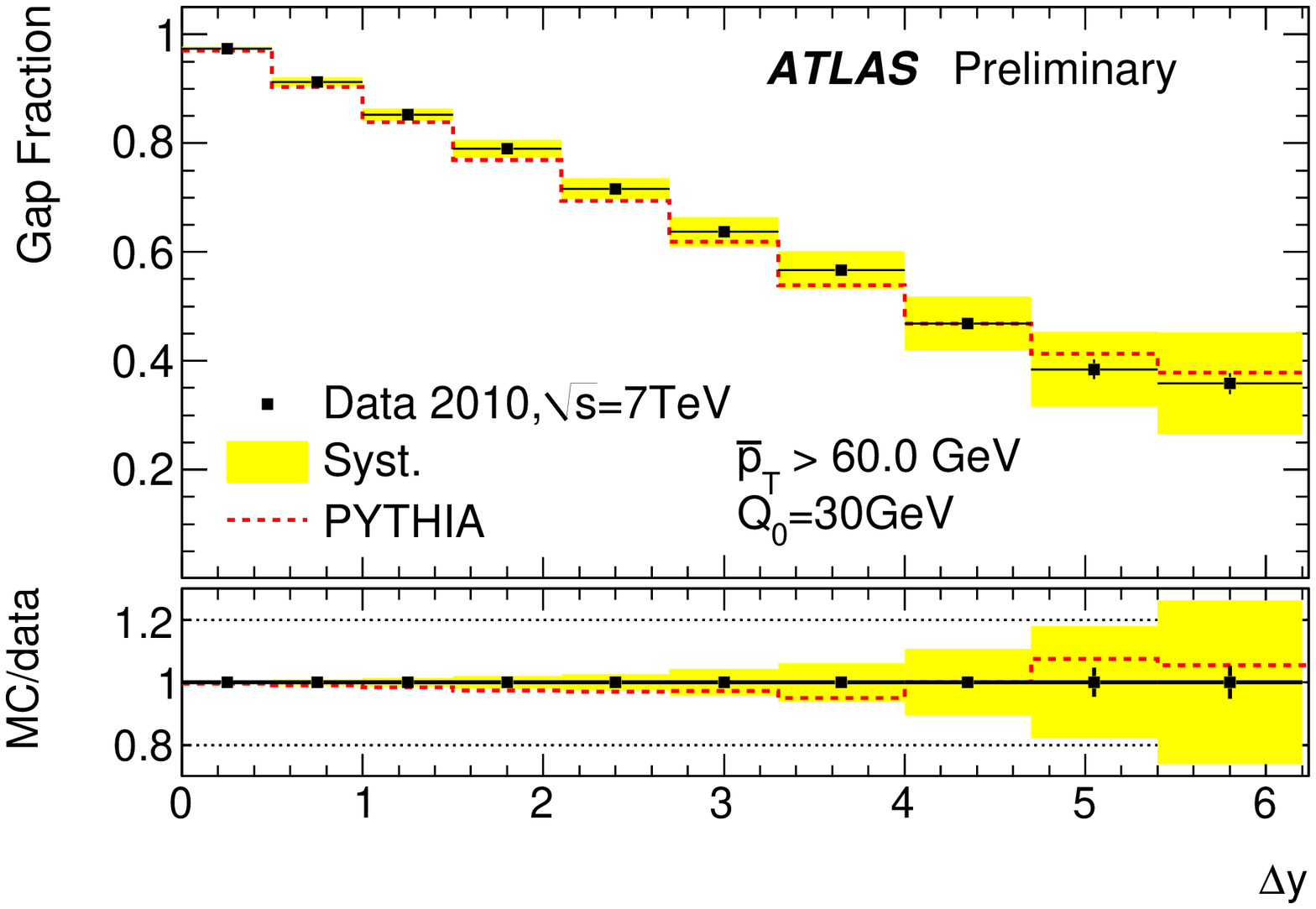}
%\label{fig:gap4}
}

\caption[]{Gap fraction for selection A is shown in (a) and (c), while gap fraction for selection B is shown in (b) and (d). The left two figures show the gap fraction as a function of the average transverse momentum, and the right two figures show the gap fraction as a function of rapidity difference between the two boundary jets.}
\label{fig:gap}
\end{figure}

It is interesting to measures the fraction of dijet events in that do not contain an additional jet in the rapidity region bounded by the dijet system.
The requirement are two good anti-$k_T$ $R=0.6$ jets with average $p_T > 60$ GeV and each with $p_T > 30$ GeV, within $|y| < 4.5$ and $\Delta y> 2$. 
Two different definitions for boundary jets are used, (A) two highest transverse momentum jets in the event, and (B) most forward and most backward jets in the event. The gap-fraction is studied as a function of avergae $p_T$ and also as a function of the rapidity separation of the jets, $\Delta $y~\cite{gap}. This can be studied to to investigate interesting QCD phenomenon, like color singlet exchange, distribution of (wide-angle) soft gluon radiation in the gap, sensitivity to BFKL-like dynamics~\cite{bfkl}.
The gap fraction is shown as a function of these two variables in \FigRef{fig:gap}. For all distributions, reasonable agreement between data and the simulation is obereved, although there is some statistical fluctuation.

%Dijet angular correlation

\begin{figure}[htbp]
\centering
\subfigure[]{
\includegraphics[scale=0.30]{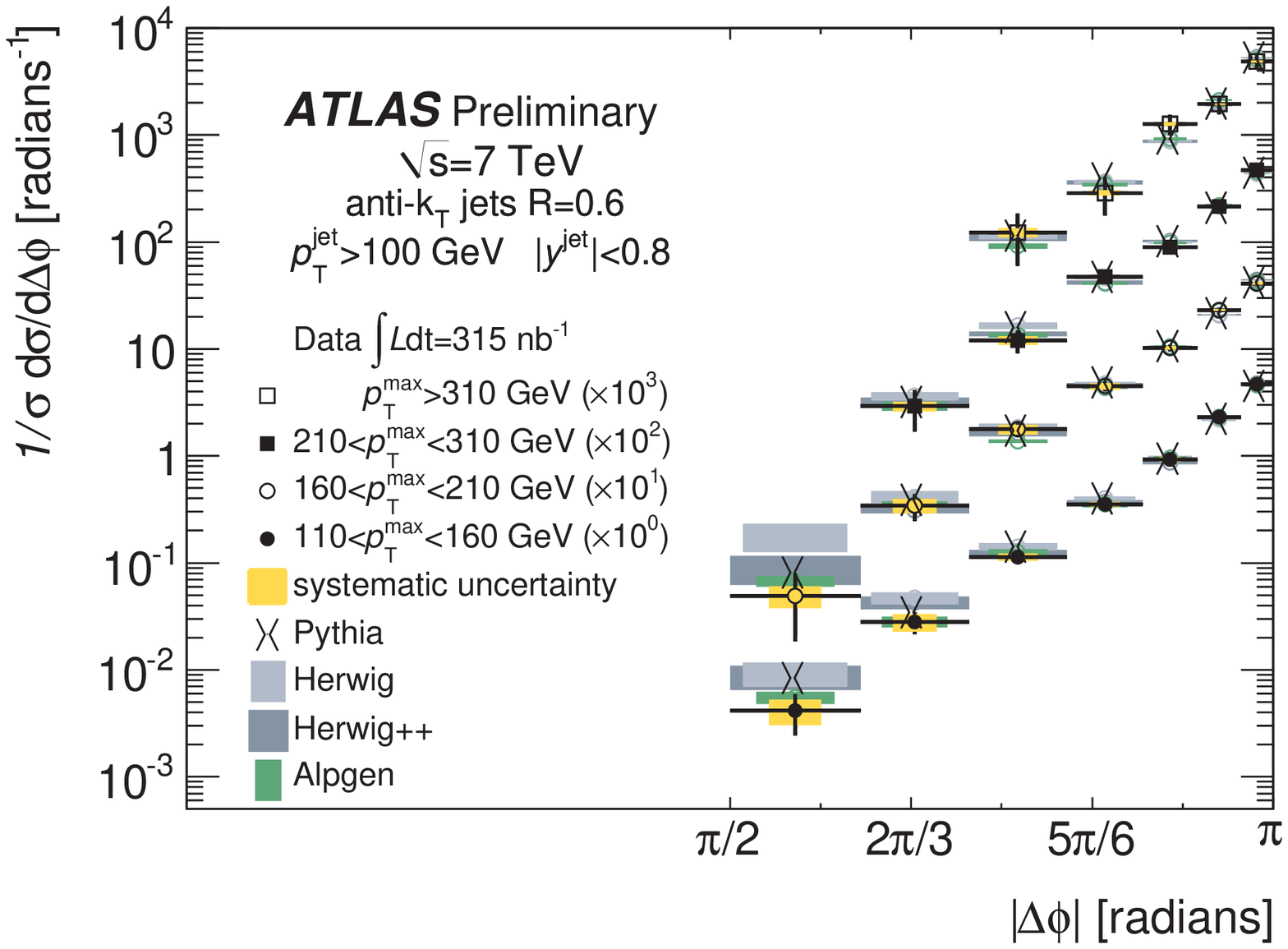}
%\label{fig:ang_lo}
}
\subfigure[]{
\includegraphics[scale=0.30]{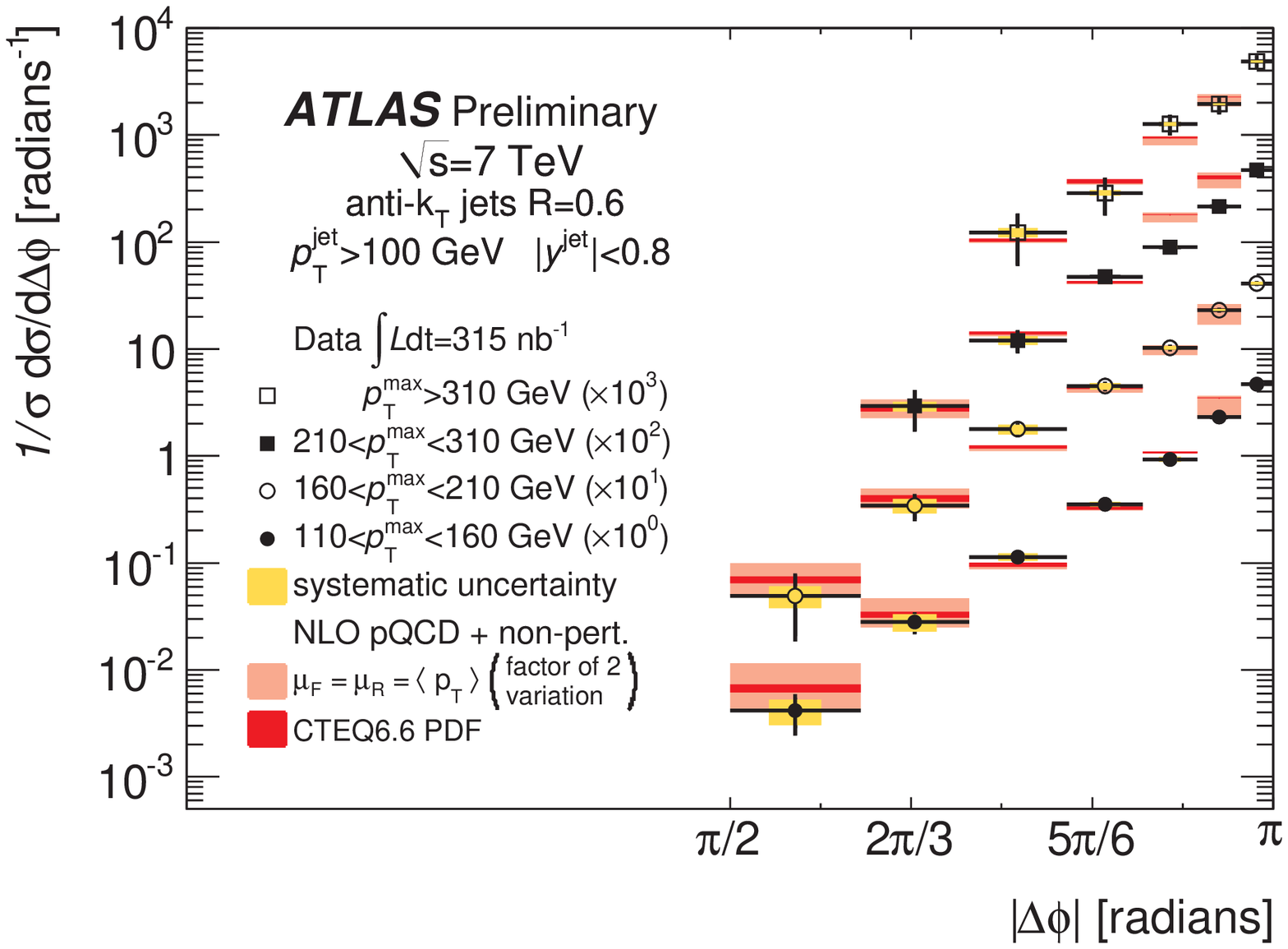}
%\label{fig:ang_nlo}
}
\caption[]{The normalized differential cross section binned in four regions based on the \pT of the leading jet is shown as a function of $\Delta \phi$ between the 2 leading jets. In (a) the data is comapred with LO MC predictions while in (b) it is compared with NLO pQCD predictions.}
\label{fig:ang}
\end{figure}

The angular difference $\Delta \phi$ between the 2 leading jets reflects the activity in the rest of the event and it is sensitive to higher order QCD radiation. Both leading jets are required to have central rapidity ($|y| < 0.8$), with jet $pT> 100$ MeV. Events are then divided into 4 bins based on leading jet \pT. The peak at $\pi$ is due the dominant final state of back-to-back dijets, and any deviation due to radiation of one or more gluons.
In \FigRef{fig:ang} the data is compared to LO and NLO predictions~\cite{ang}. Azimuthal decorrelation is sensitive to fixed order matrix elements as well as to resummation effects. There is general agreement between the MC samples and the data within the statistical uncertainties.

\section{SUMMARY}

In the first few months of LHC running, ATLAS has successfully performed different interesting QCD measurements.
Minimum bias distributions with \pT ~down to 100 MeV and the first ever underlying event analysis 
at $\sqrt{s}=$ 7 TeV are important soft QCD inputs for MC tuning. Most pre-LHC models were
seen not to agree with the these distributions, and the new tune AMBT using these measurements
is a significant improvement for minimum bias results. First measurements of kinematics for inclusive jets, dijets, and multi-jets have been performed using data at 7 TeV. The jet data are seen to be mostly consistent with NLO QCD predictions. 
ATLAS has observed first events with $p_{T,jet} \sim$ 1 TeV and with dijet mass $\sim$ 2.5 TeV, both beyond Tevatron kinematic range.
All these QCD studies are essential ingredients to `rediscover' the standard model at the LHC.

\end{document}